\documentclass[fleqn,usenatbib]{mnras}

\usepackage{newtxtext,newtxmath}

\usepackage[T1]{fontenc}
\usepackage{ae,aecompl}

\usepackage{graphicx}	
\usepackage{amsmath}	
\usepackage{amssymb}	
\usepackage[usenames,dvipsnames,svgnames,hyperref]{xcolor}
\usepackage{nccmath}
\usepackage{pdfpages}
\usepackage{float}
\usepackage{varwidth}
\usepackage{tcolorbox}
\usepackage{caption}

\usepackage{tikz}
\usetikzlibrary{positioning,calc,backgrounds}
\tikzstyle{line}=[draw]
\tikzstyle{arrow}=[draw, -latex] 

\newcommand{\upsub}[1]{\sb{\mathrm{#1}}}
\newcommand{\upsup}[1]{\sp{\mathrm{#1}}}

\def\VEL{\textsc{VELOCIraptor}}
\def\TREE{\textsc{TreeFrog}}
\def\WW{\textsc{WhereWolf}}
\def\ORB{\textsc{OrbWeaver}}

\begingroup\lccode`~=`_\lowercase{\endgroup\let~\upsub}
\begingroup\lccode`~=`^\lowercase{\endgroup\let~\upsup}

\AtBeginDocument{%
  \catcode`_=12 \catcode`^=12
  \mathcode`_="8000
  \mathcode`^="8000
}

\hypersetup{final}

\title[Extracting galaxy merger timescales I]{Extracting galaxy merger timescales I: Tracking haloes with \textsc{WhereWolf} and spinning orbits with \textsc{OrbWeaver} }

\author[R. Poulton et al.]{
Rhys J. J. Poulton,$^{1,2}$\thanks{E-mail: rhys.poulton@icrar.org}, Chris Power$^{1,2}$, Aaron S.G. Robotham$^{1,2}$ and Pascal J. Elahi$^{1,2}$
\\
$^{1}$International Centre for Radio Astronomy Research, University of Western Australia, 35 Stirling Highway, Crawley, WA 6009, Australia \\
$^{2}$ARC Centre of Excellence for All Sky Astrophysics in 3 Dimensions (ASTRO 3D)}

\date{Accepted XXX. Received YYY; in original form ZZZ}

\pubyear{2019}

\begin{document}
\label{firstpage}
\pagerange{\pageref{firstpage}--\pageref{lastpage}}
\maketitle

\begin{abstract}
Hierarchical models of structure formation predict that dark matter halo assembly histories are characterised by episodic mergers and interactions with other haloes. An accurate description of this process will provide insights into the dynamical evolution of haloes and the galaxies that reside in them. Using large cosmological $N$-body simulations, we characterise halo orbits to study the interactions between substructure haloes and their hosts, and how different evolutionary histories map to different classes of orbits. We use two new software tools - {\textsc{WhereWolf}}, which uses halo group catalogues and merger trees to ensure that haloes are tracked accurately in dense environments, and {\textsc{OrbWeaver}}, which quantifies each halo's orbital parameters. We demonstrate how \textsc{WhereWolf} improves the accuracy of halo merger trees, and we use \textsc{OrbWeaver} to quantify orbits of haloes. We assess how well analytical prescriptions for the merger timescale from the literature compare to measured merger timescales from our simulations and find that existing prescriptions perform well, provided the ratio of substructure-to-host mass is not too small. In the limit of small substructure-to-host mass ratio, we find that the prescriptions can overestimate the merger timescales substantially, such that haloes are predicted to survive well beyond the end of the simulation. This work highlights the need for a revised analytical prescription for the merger timescale that more accurately accounts for processes such as catastrophic tidal disruption.

\end{abstract}

\begin{keywords}
methods: numerical -- galaxies: evolution -- galaxies: haloes -- dark matter

\end{keywords}



\section{Introduction}

In the standard cosmological model, dark matter haloes grow hierarchically, whereby low mass haloes merge to build up progressively more massive systems. Some of these merged systems survive as substructure haloes or subhaloes \citep[e.g.][]{Tormen1997,Tormen1998,Klypin1999,Moore1999}, and can exist for long periods within their host,
undergoing many orbits \citep[e.g.][]{boylan-kolchin_dynamical_2008,jiang_n-body_2014}. The manner in which a subhalo is accreted can change drastically subsequent evolution of both the subhalo and its host halo  \citep[e.g.][]{White1978,Blumenthal1986,Dubinski1994,Mo1998}. A subhalo will continue to orbit within its host until it either (1) merges with the host, or (2) is tidally disrupted as a result of mass loss driven by tidal heating and stripping \citep[e.g.][]{Ostriker1972,Gnedin1999,Dekel2003,Taylor2004,DOnghia2010}. If the subhalo hosts a satellite galaxy but undergoes tidal disruption, then the satellite can persist even after the subhalo's disruption \citep[e.g.][]{Springel2001,Kravtsov2004,Zentner2005,Conroy2006,Natarajan2009}. The eventual fate of the satellite depends strongly on the type of orbit it was on when its host (sub)halo was lost. Galaxies on highly circular orbits can survive longer than those on radial orbits, and so can have a strong influence on the galaxy merger rate and how galaxy mergers happen \citep{Wetzel2011}. 

The merger rate depends on the dynamical friction timescale, and the dependence of this timescale on the orbit of the merging system has been modelled extensively \citep[e.g.][]{Binney1987,lacey_merger_1993,Jiang2008,boylan-kolchin_dynamical_2008,Simha2017}. These merger timescale prescriptions are used in Semi-Analytical Models (SAMs) to estimate a satellite galaxy's survival time, i.e.\ determine when it will merge with its host halo 
\citep[e.g.][]{lacey_merger_1993,Cora1997,Fujii2006,Jiang2008,boylan-kolchin_dynamical_2008,jiang_n-body_2014,Simha2017,Lagos2018}. These prescriptions are of vital
importance when coupling SAMs to large volume cosmological $N$-body simulations, which do not have either the mass or snapshot resolution to allow subhaloes to be accurately tracked to the point at which they disrupt and the galaxy merges.

To ensure that estimates for the merger timescale are accurate, high spatial and temporal resolution simulations are used
to calibrate the timescale by tracking subhaloes until they have complete disrupted and they are deemed to have merged. However, even with high resolution, tracking subhaloes is a challenging problem. In cosmological
simulations, (sub)haloes can disappear and then re-appear, sometimes multiple times, in consecutive halo 
catalogues, especially when they are in dense environments, which can lead to estimates of
premature merging and artificially reduced merger timescales \citep{Poulton2018,Elahi2019a}. 

In this work, we present two new open source software tools - \textsc{WhereWolf}\footnote{https://github.com/rhyspoulton/WhereWolf} and \textsc{OrbWeaver}\footnote{https://github.com/rhyspoulton/OrbWeaver}.
\textsc{WhereWolf} is a \emph{(sub)halo ghosting} tool, which is used to track (sub)haloes 
even after they have been lost by the halo finder, and to supplement halo catalogues with these recovered (sub)haloes. 
We show how using \textsc{WhereWolf} can improve measurements of the subhalo/halo mass function, and present estimates of the distribution of radii at which subhaloes merge. \textsc{OrbWeaver}, extracts orbital properties, such as eccentricity and orbital energy, at key points in a subhalo's orbit. We show the distribution of orbital properties recovered by \textsc{OrbWeaver},
and we compare our results to previous work. Finally, we show how current prescriptions of the merger timescale from \citealt{Binney1987,lacey_merger_1993,Jiang2008,boylan-kolchin_dynamical_2008} perform on the latest generation of large N-body simulations.

This paper is organised as follows: in \S2 we discuss the data and the codes used; our results on
constructing halo orbits are presented in \S3; orbital analysis of (sub)haloes is presented in \S4; and in \S5 we conclude
the paper with a discussion on the importance of accurate halo tracking in simulations.

\section{Input Catalogues}

The simulations used in this work come from two suites of $N$-body simulations; (1) \textsc{genesis} (Elahi et at., in preparation), with volumes ranging from 26.25 to 500 Mpc $h^{-1}$ and between 324$^3$ to 4320$^3$ particles; and (2) \textsc{SURFS } (Synthetic UniveRses for Future Surveys) \citep{elahi_surfs:_2018}, with volumes ranging from 40 to 900 Mpc $h^{-1}$ and between 512$^{3}$ to 2048$^3$ particles.
Both suites of simulations were run assuming a $\Lambda$CDM Planck 2015 cosmology with $\Omega_{\mathrm{M}}$ = 0.3121, $\Omega_{\mathrm{b}}$ = 0.6879, $\Omega_{\Lambda}$ = 0.6879, a normalisation $\sigma_{\mathrm{8}}$  = 0.815, a primordial 
spectral index $n_{\mathrm{s}}$ = 0.9653, and a $H_0=67.51 {\rm km\,s}^{-1},{\rm Mpc}^{-1}$ \citep{planck_collaboration_planck_2015}. A total 200 snapshots are stored, evenly spaced in logarithmic expansion factor ($a=1/(1+z)$) between $z = 24$ to $z = 0$. This high cadence enables an accurate capturing of the evolution of dark matter haloes and their orbits. Both suites of simulations have been run with a memory-lean version of GADGET2 \citep{springel_simulations_2005}

For this study, we focus on the \textsc{genesis} simulation, with a box size of 105 Mpc $h^{-1}$ and 2048$^3$ particles, implying a particle mass of 1.73 $\times 10^{7}$ M$_{\odot}$; and the \textsc{SURFS} simulation, with a box size of 40 Mpc $h^{-1}$ and 512$^3$particles, with a particle mass of 4.12 $\times 10^7$ M$_{\odot}$. These boxes provide us with a statistical sample of well resolved central haloes, with virial masses $\sim$ 10$^{10.5}$M$_{\odot}$; at least 1,000 particles for centrals, and at least 20 particles for subhaloes. Unless otherwise stated, we use \textsc{genesis} volume for our analysis.

Halo catalogues are constructed using \VEL\, a 6-Dimensional Friends-of-Friends (6D-FoF) phase space halo finder \citep{elahi_peaks_2011,elahi_streams_2013,Canas2019,Elahi2019b}, while trees are constructed using \TREE\ \citep{Elahi2019a}, which is a particle correlator that can link across multiple snapshots and halo catalogues. Importantly, \TREE's ability to link across multiple snapshots is vital for tracking subhaloes as they orbit within highly overdense regions. While a subhalo may not be present in a pair of halo catalogues at consecutive output times, it may be present in halo catalogues at a later time, and so there might be gaps in the subhalo's history. This has led to the development of the halo tracking tool known as \WW.

\subsection{\WW}

\begin{figure}
\centering
\begin{tikzpicture}[
        roundnode/.style={rectangle, rounded corners = 5pt, draw=ForestGreen, very thick, text width=0.16\textwidth,align=center},
    squarednode/.style={rectangle, draw=Cerulean, very thick, minimum size=5mm, text width=0.16\textwidth, align=center},
    node distance=5mm
    ]

    \node[squarednode](search){
    [1]: Search for (sub)haloes that have been lost and can be tracked
    };
    
    \node[roundnode](calc)[right=of search]{
    [9]: Use most bound particles (MBPs) as reference position and velocity
    };
    
    \node[roundnode](extract)[below=of search]{
    [2]: In the final snapshot that the (sub)halo is identified, extract particle data
    };
    
    \node[roundnode](Pos)[right=of extract]{
    [10]: Use the previous M$_{vir}$ and R$_{vir}$ to calculate V$_{\rm esc}$
    };
    \path (extract) -- (Pos) coordinate[midway] (CENTER);
    \node[roundnode](bound)[below=12mm of CENTER]{
    [3]: Calculate bound mass from the particles with r < R$_{\rm vir}$ and V$_{\rm r}$ < V$_{\rm esc}$ with reference to the (sub)haloes.
    };
    
     \node[squarednode](Add)[right= of bound]{
    [8]: Add the candidate (sub)halo to the halo catalogue.
    };
    
    \node[roundnode](MBP)[below= of bound]{
    [4]: Find the 10\% or 10 MBPs by using equation \ref{equ:mbp}.
    };
    
    \node[roundnode](host)[right= of MBP]{
    [7]: Find a if the candidate (sub)halo has a host or not.
    };
    
    \node[roundnode](Cond)[below= of MBP]{
    [5]: Check if candidate has more than 20 bound particles
    };
    \node[roundnode](Match)[right= of Cond]{
    [6]: If not "filling in the gap" then check if there is a matching (sub)halo
    };
    
    \node[squarednode](Stop)[below= of Cond]{
    [11]: Stop tracking
    };
    
    \draw[thick,->] (search.south) -- (extract.north);
    \draw[thick,->] (calc.south) -- (Pos.north);
    \draw[thick,->] (extract.south) |- ++(0,-2mm) -| (bound.north);
    \draw[thick,->] (Pos.south) |- ++(0,-3.9mm) -| (bound.north);
    \draw[thick,->] (bound.south) -- (MBP.north);
    \draw[thick,->] (MBP.south) -- (Cond.north);
    \draw[thick,->] (Cond.south) -- node[anchor=east]{No} (Stop.north);
    \draw[thick,->] (Cond.east) -- node[anchor=south]{Yes} (Match.west);
    \draw[thick,->] (Match.south) |- node[anchor=west, pos=0.25]{Yes} (Stop.east);
    \draw[thick,->] (Match.north)  -- node[anchor=west]{No}  (host.south);
    \draw[thick,->] (host.north)  -- (Add.south);
    \draw[thick,dashed,->] (Add.north)   |- node[anchor=west,pos=0.4,rotate=-90,yshift=0.2cm]{Next snapshot} (calc.east);
\end{tikzpicture}
\caption{The activity chart of \WW\ . }
\label{fig:WWflow}
\end{figure}
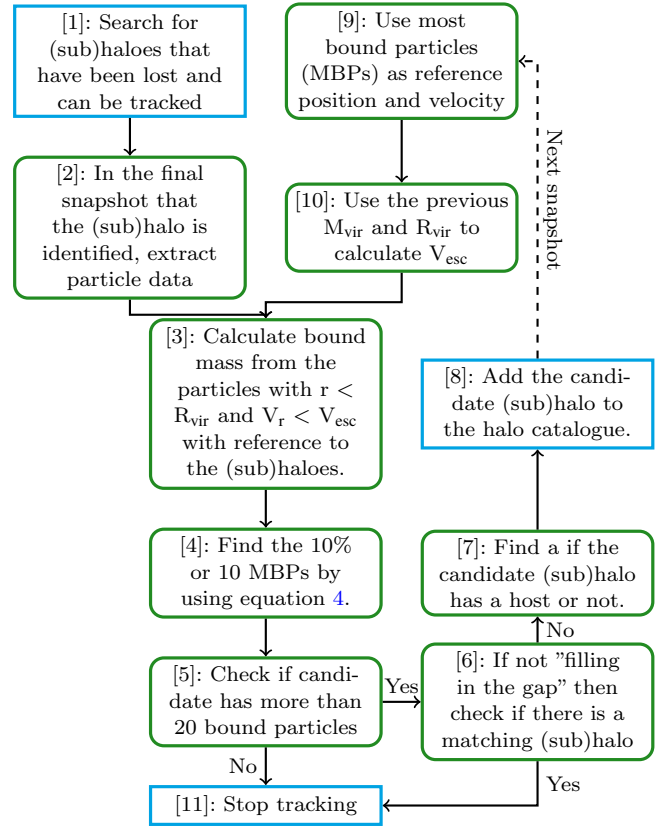

\begin{figure}
\centering
\includegraphics[width=0.47\textwidth]{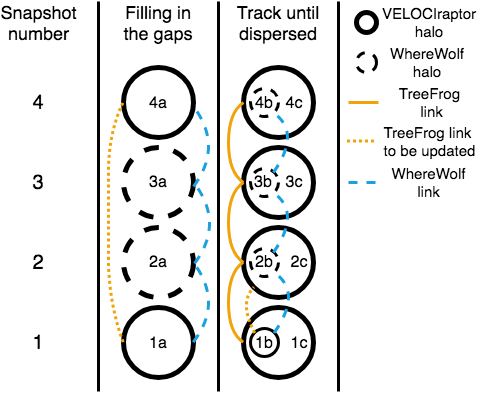}
\caption{A schematic diagram illustrating the two different ways in which \WW\ tracks a (sub)halo. The leftmost column shows snapshot number, with an increasing number indicating increasing time in the simulation. The middle two columns show the cases in which \WW\ inserts ``missing" (sub)haloes in each snapshot (``Filling in the gaps"), or tracks a (sub)halo until it becomes unbound (``Track until dispersed"). The rightmost column provides the key.
\label{fig:trackcond}}
\end{figure}

\WW\ is a (sub)halo tracking tool, originally introduced in \citealt{Poulton2018}. A representation of the \WW\ decision tree is shown in Figure \ref{fig:WWflow}. In summary, (sub)haloes are identified by gaps in \TREE\ trees, and their particles are extracted from a \VEL\ catalogue and propagated forward in time to see if these (sub)haloes remain bound at later times, or if they have been permanently disrupted. There are two cases in which \WW\ is triggered:

\medskip

\noindent 1. a (sub)halo has a descendant that is more than one snapshot away, which can occur during \TREE\ multi-snapshot linking;

\smallskip

\noindent 2. two (sub)haloes merge, and a (sub)halo's descendant becomes ambiguous.

\medskip

A schematic of the above two cases is shown in Figure \ref{fig:trackcond}. In both cases, \WW\ tracks a (sub)halo if N$_{part}$ > N$_{min,track}$, where N$_{min,track}$ is the minimum particle number. For this paper, N$_{min,track}$ is set to 50 particles because this is above the 20 particle limit for a (sub)halo to exist in the catalogue and to be tracked for at least one snapshot (boxes 1 and 2 in Figure \ref{fig:WWflow}). 

\bigskip

\subsubsection{\WW\ in Depth}

In this Section we explain how \WW\ works in detail.

\medskip

\noindent \textbf{1. Boundedness calculation.}
\WW\ estimates the boundedness of a (sub)halo by calculating the escape velocity (V$_{ esc}$) 
profile, which can be related to the circular velocity by: $V_{ esc}(r)   =  \sqrt{2} V_{ circ}(r)$. Assuming a Navarro Frenk and White profile\footnote{We use a theoretical model to calculate Vcirc because direct measurement from the simulation can be noisy, especially at small radii where there is little enclosed mass. Also, because $V_{circ}$ is being calculated for subhaloes, the particles at large radii are also subject to the effects of the host.  Hence by using the theoretical model, we can use more stable quantities e.g. Mvir, Rvir and c. Here the concentration is evaluated by assuming the last value that was measured by \VEL. } \citep[NFW;][]{navarro_universal_1997}, V$_{ circ}$ is given by:

\begin{ceqn}
\begin{equation}
V_{ circ}(r)   = V_{ vir} \sqrt{\frac{f(cx)}{f(x)} };
\label{equ:Vcirc}
\end{equation}
\end{ceqn}
here $c$=$R_{ s}/R_{ vir}$ is the concentration, where $R_{ s}$ is the scale radius and $R_{ vir}$ is the virial radius, defined such the mean interior density $\bar{\rho}$ = 200$\rho_{crit}$. The function $f(x)$ is given by
\begin{ceqn}
\begin{equation}
f(x) = \mathrm{ln}(1+x) + \frac{x}{1+x},
\end{equation}
\end{ceqn}
with $x$ = $r/R_{ vir}$. V$_{ vir}$ is calculated as,
\begin{ceqn}
\begin{equation}
V_{ vir}   =  \sqrt{\frac{GM_{ vir}}{R_{ vir}}},
\label{equ:Vvir}
\end{equation}
\end{ceqn}
where $G$ is the gravitational constant in relevant units, and $M_{ vir}$ is the mass of the (sub)halo contained within $R_{ vir}$ (box 3 in Figure \ref{fig:WWflow}). Particles are considered bound if they are below V$_{esc}$ and are with R$_{vir}$.

\begin{figure}
\includegraphics[width=0.47\textwidth]{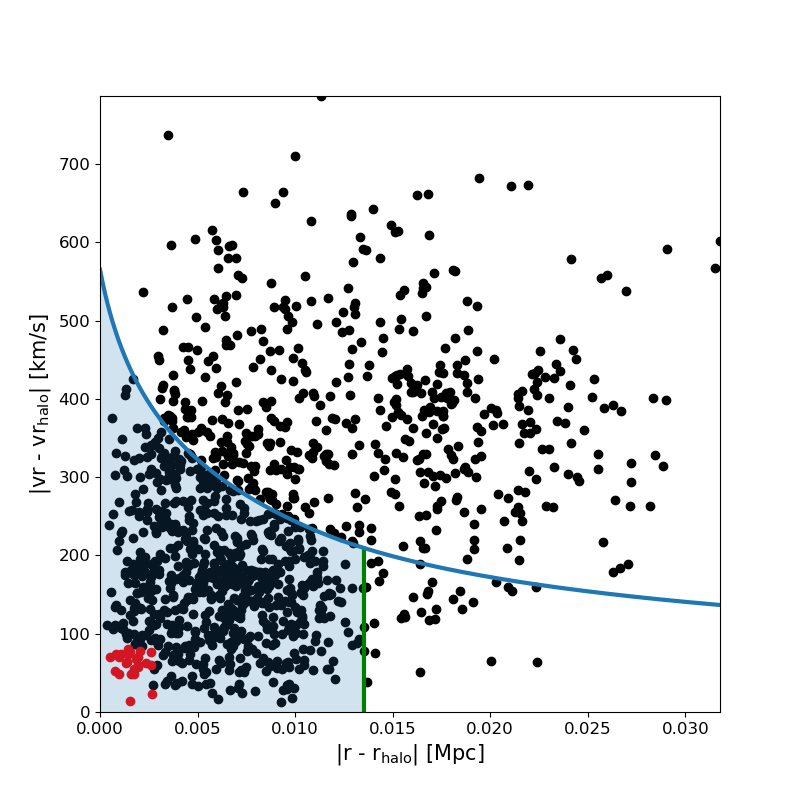}
\caption{Phase-space distribution of all subhaloes particles that are with its 6D-FoF, shown one snapshot after the initial tracking snapshot. The figure shows how the (sub)halo particle's halo-centric velocities vary with halo-centric radius; the blue curve shows the caustic defined by $V_{ esc}$ while the green vertical line indicates the radial boundary of the (sub)halo. The red points are the 10\% MBPs described below.\label{fig:caustic}}
\end{figure}

\medskip

\noindent \textbf{2. Tagging candidate tracked (sub)halo particles.}
In the initial snapshot at which tracking begins, $M_{ vir}$ and $R_{ vir}$ are taken from the \VEL\ catalogue. From this, \WW\ determines whether or not a particle is bound to the (sub)halo using the above binding conditions. An example candidate (sub)halo is shown in Figure \ref{fig:caustic}, where the shaded region indicates which particles are bound to it. The most bound particles (MBPs) are determined by calculating the following distance metric for all particles:
\begin{ceqn}
\begin{equation}
D^2 \ = \ \frac{r^2}{R_{vir}^2} \ + \ \frac{v_{r}^2}{V_{esc}^2}
\label{equ:mbp}
\end{equation}
\end{ceqn}
where $r$ and $v_{r}$ are the relative position and velocity of the (sub)halo to its centre of mass. This metric is used to find the MBPs, where the minimum 10\% are defined as the MBPs, with a minimum of 10 MBPs (box 4 in figure \ref{fig:WWflow}).

\medskip

\noindent \textbf{3. Following tracked (sub)halo particles.}
The MBPs in the example shown in Figure \ref{fig:caustic} are indicated by red points; these populate the core of the caustic and are used to determine reference positions and velocities for the tracked (sub)halo at the next snapshot (box 9 in Figure \ref{fig:WWflow}). \WW\ estimates $R_{ vir}$ and $M_{ vir}$ from the MBPs by finding the radius at which the enclosed density drops below 200$\rho_{crit}$.

At this point, $V_{ esc}$ is re-calculated to find which particles belong to the (sub)halo (boxes 10 and 3 in Figure \ref{fig:WWflow}), along with re-calculating the MBPs (box 4 in Figure \ref{fig:WWflow}). Only particles within 10 times the mean radius are retained; this is because particles can move large distances between snapshots and become completely unbound from the (sub)halo. If there are fewer than 20 bound particles, then the (sub)halo is not tracked forward to the next snapshot (boxes 5 and 11 in Figure \ref{fig:WWflow}), otherwise, an attempt is made to identify the (sub)halo in the \VEL\ catalogue  (box 6 in Figure \ref{fig:WWflow}). This is done using a \textsc{K-D tree} \citep{bentley_multidimensional_1975} to find (sub)haloes that fall within the tracked (sub)halo's $R_{ vir}$, and a (sub)halo is matched to the tracked (sub)halo if and only if it does not have a progenitor, i.e. the (sub)halo's branch is truncated. This reduces the number of truncated branches and avoids terminating a branch when it does not have a descendant, which would happen when a non-truncated branch is matched. In this situation the \WW\ connection would overwrite the \TREE\ connection, meaning that a (sub)halo that was previously connected would not have a descendant.

Matching is achieved by comparing particles that belong to each (sub)halo and using a merit function\footnote{For details on the full merit functions that are used, please see \citep{Elahi2019a}.}, such as,
\begin{ceqn}
\begin{equation}
M   = \frac{N_{ sh}^2}{N_1 N_2};
\label{equ:merit}
\end{equation}
\end{ceqn}
here $N_{ sh}$ is the number of particles shared between two (sub)haloes; $N_{1}$ is the number of particles in the \WW\ (sub)halo; and $N_{2}$ is the number of particles in the matched \VEL\ (sub)halo. A match occurs when $M$>0.025, and the descendant for the tracked (sub)halo points to the matched \VEL\ (sub)halo. We can also apply the same merit to the MBPs using the same threshold. 

If a match is not found, (sub)haloes that are within the tracked (sub)halo's $R_{ vir}$ are used to identify if the \WW\ (sub)halo stays within 0.1 R$_{ vir}$ of a \VEL\ (sub)halo for more than 3 snapshots; if so, the \WW\ (sub)halo is considered completely merged with the \VEL\ (sub)halo. This removes any (sub)haloes that remain bound, even after they have merged formally, and avoids (sub)haloes persisting for many snapshots (box 6 in Figure \ref{fig:WWflow}). 

If no match is found and the tracked (sub)halo has not remained within 0.1 R$_{vir}$ of a \VEL\ (sub)halo then the tracked (sub)halo is "accepted" and \WW\ determines if the tracked (sub)halo has a host. The host of a \WW\ (sub)halo is determined by checking if the initial \VEL\ (sub)halo (which was used to identify the \WW\ (sub)halo) has a host. If a host exists, then the \WW\ (sub)halo is set as a descendant if it exists in the next snapshot. If there is no host for the initial \VEL\ (sub)halo or if it does not exist in the next snapshot, a search is carried out to determine if the (sub)halo has a new host. This is done by 
checking if the \WW\ (sub)halo lies within the R$_{ vir}$ of a \VEL\ halo using the \textsc{K-D tree}, and then checking if it is bound to the
halo by evaluating

\begin{ceqn}
\begin{equation}
0.5 M_{ WW} V_{ WW} - \frac{G M_{ VEL} M_{ WW}}{|r_{ WW} - r_{ VEL}|} < 0;
\label{equ:bound2halo}
\end{equation}
\end{ceqn}

where $M_{ WW}$ is the virial mass of the \WW\ (sub)halo; $V_{ WW}$ is the velocity of the \WW\ (sub)halo relative to the \VEL\ halo; $M_{ VEL}$ is the virial velocity of the \VEL\ halo; and |r$_{ WW}$ - r$_{ VEL}$| is the separation between the \WW\ and \VEL\ haloes. Equation \ref{equ:bound2halo} establishes whether or not the \WW\ halo is bound to the \VEL\ halo (box 7 in Figure \ref{fig:WWflow}). If the \WW\ (sub)halo can be tracked, it is added to the \VEL\ halo catalogue (box 8 in Figure \ref{fig:WWflow}).

It is clear that adding in these (sub)haloes will impact the reconstruction of subhalo orbits. We can study the impact of \WW\ on our orbital reconstruction using our newly developed orbital analysis tool, which is known as \ORB. This is introduced and discussed in detail in the next Section.

\subsection{\ORB}

\begin{figure}
\centering
\begin{tikzpicture}[
    roundnode/.style={rectangle, rounded corners = 5pt, draw=ForestGreen, very thick, text width=0.3\textwidth,align=center},
    squarednode/.style={rectangle, draw=Cerulean, very thick, minimum size=5mm, text width=0.3\textwidth, align=center},
    node distance=5mm
    ]
    \node[squarednode](searching){
   [1]: Find host of interest from the halo catalog
    };
    \node[roundnode](fb)[below=of searching]{
    [2]: Identify any satellites that come within N $\times$ R$_{\rm vir}$ of the host of interest across all cosmic time
    };
    \node[roundnode](outliers)[below=of fb]{
    [3]: Extract the full history of the satellite relative to host 
    };
    \node[roundnode](link)[below=of outliers]{
    [4]: Give the satellite a unique orbitID to identify its orbit
    };
    \node[roundnode](process)[below=of link]{
    [5]: Process the satellite relative to its host identifying key points in its orbit and interpolate to find the exact properties at each point
    };
    \node[roundnode](clean)[below=of process]{
    [6]: Clean the (sub)halo's orbit removing any wobbles in its orbit
    };
    \node[squarednode](newsearch)[below=of clean]{
    [7] Output the key points to a file.
    };
    
    \draw[thick,->] (searching.south) -- (fb.north);
    \draw[->,thick] (fb.south) -- (outliers.north);
    \draw[->,thick] (outliers.south) -- (link.north);
    \draw[->,thick] (link.south) -- (process.north);
    \draw[->,thick] (process.south) -- (clean.north);
    \draw[->,thick] (clean.south) -- (newsearch.north);
    
\end{tikzpicture}
\caption{The activity chart of \ORB\ . }
\label{fig:ORBflow}
\end{figure}
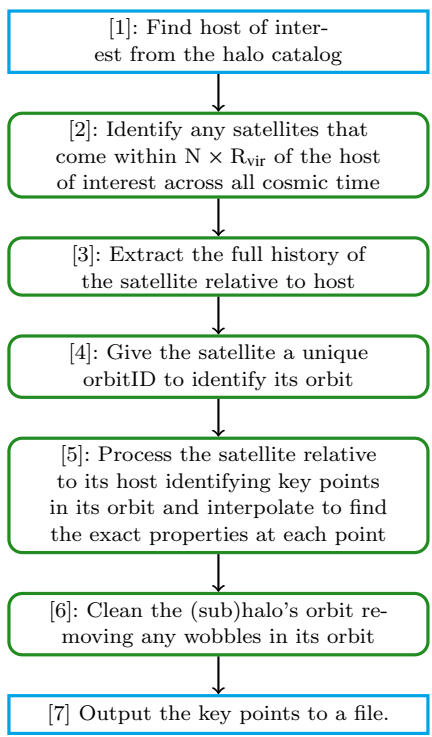

\ORB\ is a tool for processing merger trees and extracting orbital catalogues for a statistical sample of (sub)haloes from cosmological simulations. A representation of how \ORB\ works is shown in Figure \ref{fig:ORBflow}. In summary, for each halo in a \VEL\ catalogue, \ORB\ tracks the full histories of all other (sub)haloes that have passed within N $\times$ R$_{ vir}$  (where N is a positive integer) of the host of interest (R$_{vir,host}$) across cosmic time (boxes 1, 2 and 3 in Figure \ref{fig:ORBflow}).  

Once processed, (sub)haloes are assigned {\small{OrbitIDs}}\footnote{A given (sub)halo can have multiple {\small{OrbitIDs}}.} and written to catalogue (box 4 in Figure \ref{fig:ORBflow}). In this way, every (sub)halo that satisfies a given {\small{OrbitIDs}} can be extracted, and the orbital properties -- a list of which is given in Appendix \ref{app:orbdesc} -- of this set are recorded at two points:
\begin{itemize}
\item the \textbf{crossing point}, at which the (sub)halo crosses $M$ times the host virial radius (R$_{ vir,host }$) with $0.0  < M \leq 4.0$; and
\item the \textbf{apsis point}, at which a (sub)halo's radial velocity vector (\textbf{V}$_{ rad}$) changes direction. If the radial velocity goes from negative to positive, it is a peri-centric apsis, otherwise it is an apo-centric apsis.
\end{itemize}
A root finding algorithm is used to find the crossing and apsis points so that (sub)halo properties -- position, velocity, mass, maximum circular velocity (V$_{ max}$), and virial radius -- are interpolated to the exact point that the event happens (box 5 in Figure \ref{fig:ORBflow}). Most orbital properties are calculated at both points, but there are some properties (such as eccentricity ($\epsilon$)) that are calculated only for apsis points. The table in Appendix \ref{app:orbdesc} explicitly states when each property is computed.

\subsubsection{Orbit cleaning} \label{sec:cleaning}

\begin{figure}
\centering
\includegraphics[width=0.47\textwidth]{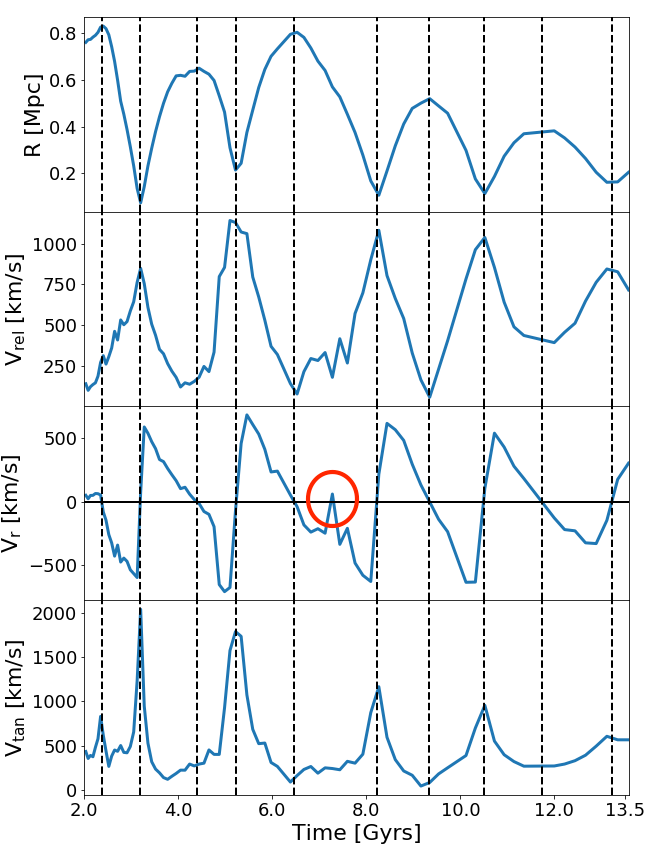}
\caption{ From top to bottom, this Figure shows the radius of the orbiting (sub)halo with respect to its host, the relative velocity between the host and the orbiting (sub)halo, the radial velocity of the orbiting (sub)halo with respect to its host and the tangential velocity of the orbiting (sub)halo around its host all as a function of time. The vertical dashed lines show the Apsis points of the orbit, and the circle shows a {\it wobble} in the orbit that could have generated a spurious apsis point. This plot is shown from the time when the orbiting (sub)halo entered 3 R$_{vir,host}$ \label{fig:cleanorb}}
\end{figure}

Once orbits have been processed they are cleaned to remove {\emph{duplicate crossing points}} or {\emph{spurious apsis points}} (box 6 in Figure \ref{fig:ORBflow}). 
\begin{itemize}
\item {\emph{Duplicate crossing points}} arise because of fluctuations in R$_{ vir,host}$. The orbiting (sub)halo can be within the R$_{ vir,host}$ in one snapshot, but outside in the
next because R$_{ vir, host}$ has shrunk, and so it appears to have crossed R$_{ vir, host}$ twice. These duplicate crossing points are removed by storing only the first instance of the infall crossing point and only storing further crossing point once the (sub)halo has had a least one outgoing crossing point. This means that the orbiting (sub)halo is set to cross R$_{ vir, host}$ only once, at first infall, and must have an outgoing crossing point before it {\it infalls} again.

\item {\emph{Spurious apsis points}} arise because the orbiting (sub)halo can interact with other orbiting (sub)haloes, which can cause perturbations in its orbit; it can also occur when the (sub)halo is part of a larger, infalling, group. In both cases, the orbiting (sub)halo can experience large changes in \textbf{V}$_{ rad}$ as well as brief changes in its direction, which can lead to it being classified as having had a apsis point about the halo it is orbiting. 

\end{itemize}
An example spurious apsis point is shown in Figure \ref{fig:cleanorb}, which shows properties of a (sub)halo as it orbits its host. The red circle highlights the occurrence of a incorrectly identified apsis point based on when the radial velocity changes direction. This is caused by the large change in the relative velocity (V$_{ rel}$), which occurs because of the orbiting (sub)halo gravitationally interacting with another, similar mass, subhalo at this snapshot. This interaction is cleaned by discarding any apsides that happen within two snapshots, as the orbit will not have been adequately sampled.

\section{Results}

\subsection{{\small{WhereWolf}}: Halo Mass Functions \& Merging Radii, R\texorpdfstring{$_{merge}$}{merge}}

\begin{figure*}
\centering
\includegraphics[width=0.9\textwidth]{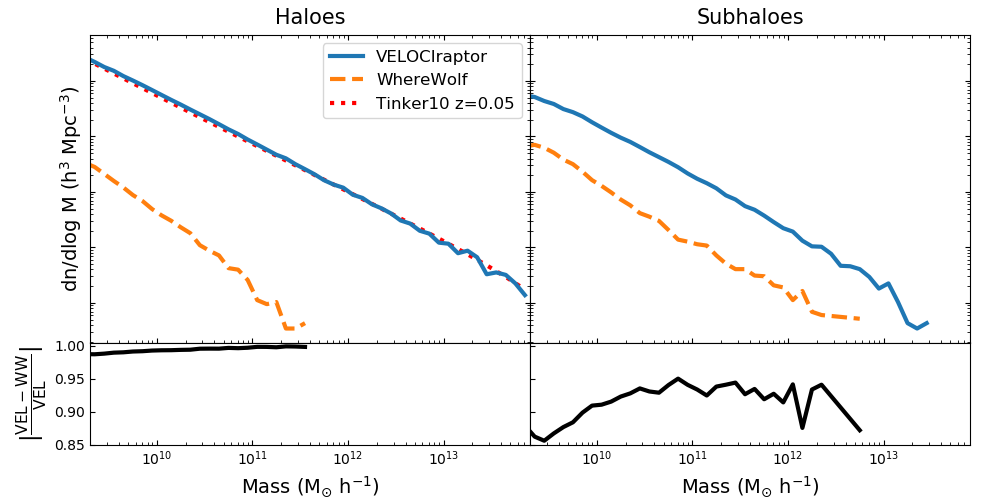}
\caption{These plots show the halo mass function (HMF, top left panel) and the subhalo mass function (SHMF, top right panel) found by \WW\ (dashed, orange lines) and from the \VEL\ catalogue (solid, blue lines) for the 105 Mpc $h^{-1}$ simulation with 2048$^3$ particles. These have been shown at z=0.05, so mass functions include any (sub)haloes that have been inserted by \WW\ for the "gap filling" events. We also plot the halo mass function from \citep{Tinker2010} on the left panel, calculated using \textsc{hmfcalc} \citep{Murray2013} for comparison.  The bottom panels show the relative difference between the \WW\ and \VEL\ mass functions. \label{fig:hmf}}
\end{figure*}

We begin by showing halo mass functions (HMF) and subhalo\footnote{By subhalo here we mean a halo that is not the top of its spatial hierarchy} mass functions (SHMF) in Figure \ref{fig:hmf}, where we have used the 105$^3$ Mpc $h^{-1}$ box with 2048$^3$ particles.  The top left panel shows the HMF , while the top right panel shows the SHMF for \WW\ (dashed orange line) and \VEL\ (solid blue line). Most of the subhaloes and haloes populated by \WW\ are at masses $\lesssim 10^{12} M_{\odot} h^{-1}$ and $\lesssim 10^{11} M_{\odot} h^{-1}$ respectively, which reflects the greater likelihood that lower mass (sub)haloes are more likely to be lost by \VEL\ and to be picked up by  \WW\'s tracking. The power-law slopes of the respective HMF and SHMF at these low masses are in reasonable agreement. The bottom panel of \ref{fig:hmf}, shows the relative difference between the catalogues HMF (solid line) and SHMF (dashed line). This demonstrates that \WW\ identifies an increased number of subhaloes in the catalogue, where the difference can be as high as 15\% in some mass bins with an average increase of 10\% more subhaloes than \VEL. In comparison, \WW\ does not track many haloes whereby the increase in the number of haloes is  <1\% . This shows that the integration of the \WW\ haloes into the \VEL\ catalogues does not have an effect on the shape and amplitude of the HMF derived from the \VEL\ catalogue, which is shown to be in good agreement with various HMFs such as \citep{Tinker2010} shown in the top panel of Figure \ref{fig:hmf} (see \citep{elahi_surfs:_2018} for other mass functions). 

\begin{figure}
\centering
\includegraphics[width=0.47\textwidth]{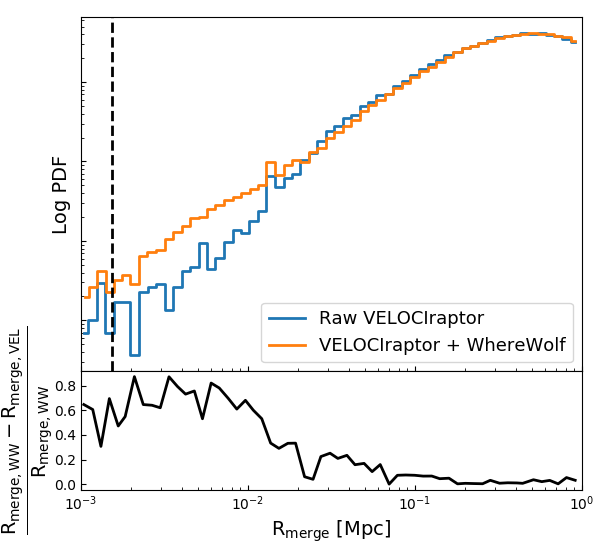}
\caption{The top panel shows the log probability density function (PDF) of the radius at which subhaloes merge, for \VEL\ (blue) and \VEL\ + \WW\ (orange); the black dashed line represents the softening length of the simulation. The lower panel shows the relative difference between the two curves shown in the top panel.  \label{fig:RmergeComp}}
\end{figure}

In Figure \ref{fig:RmergeComp} we show the probability distribution function (PDF) of the radii at which subhaloes merge with their hosts (R$_{merge}$), estimated from the original \VEL\ catalogues (heavy dark solid histogram) and the revised \VEL\ + \WW\ catalogues (heavy light solid histogram). This demonstrates that \WW\ tracks subhaloes to smaller radii before they merge, while also extending the temporal baseline over which we can track subhaloes and characterise their orbital properties. This highlights that \WW\ improves tracking of lower mass haloes and subhaloes, which are the most susceptible to being lost in any halo finding algorithm, and allows their orbits to tracked for longer and into higher overdensity regions.

\subsection{\ORB: Orbital Parameters}

\begin{figure}
\centering
\includegraphics[width=0.45\textwidth]{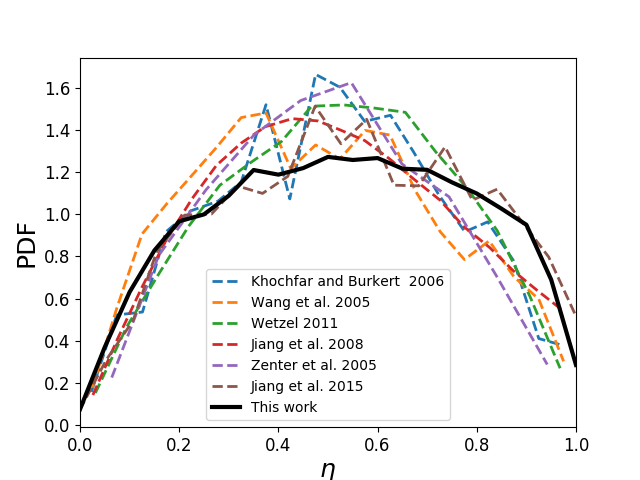}
\caption{The distribution of $\eta$, compared to previous works. \label{fig:epsilonPDF} }
\end{figure}

\ORB\ computes a multitude of properties that characterise the orbits of subhaloes around their host haloes. For this paper we focus on a few key properties, such as circularity ($\eta$) defined as

\begin{ceqn}
\begin{equation}
    \eta \ = \ \frac{J_{halo}}{J_{circ}(E)},
\end{equation}
\end{ceqn}
where $J_{halo}$ is the specific angular momentum of the orbiting (sub)halo and $J_{circ}(E)$ is the specific angular momentum of the equivalent circular orbit with the same energy (E); both of these quantities are calculated at first infall (i.e.\ when the subhalo first crosses R$_{vir,host}$). The distribution of $\eta$ is shown in Figure \ref{fig:epsilonPDF} for all (sub)haloes at $z=0$ that had at least 1,000 particles at infall, corresponding to a mass of 1.5 $\times$ $10^{10}$M$_{\odot}$. This distribution is broad and peaks at $\eta = 0.52$, which is in good agreement with previous work \citep{Tormen1997,Zentner2005,Wang2005,Khochfar2006,Jiang2008,Wetzel2011,Jiang2015,vandenBosch2017}, and shows that most satellite orbits are neither preferentially circular nor highly radial. 

\begin{figure}
\centering
\includegraphics[width=0.5\textwidth]{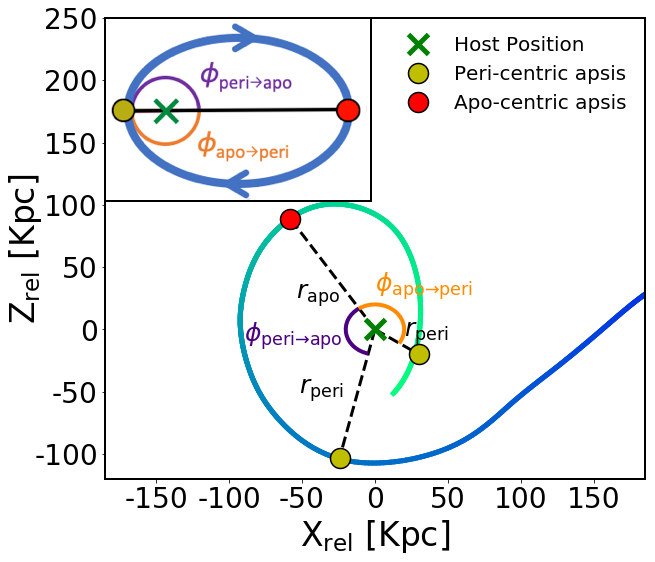}
\caption{An example planar orbit on the $x-z$ plane, showing how $\epsilon$ and $\phi$ are calculated. The lines represent the simulation outputs, where the darker blue is early times and light green is late times. This figure shows where $\phi_{peri \to apo}$ and $\phi_{apo \to peri}$ are calculated, along with a simple schematic in the top left corner. \label{fig:example}}
\end{figure}

We also study eccentricity of the orbit ($\epsilon$) and angle subtended since last apsis of an orbiting subhalo relative to the host halo ($\phi$). Here $\epsilon$\footnote{We refer the reader to Appendix \ref{app:compEccDef} for discussion about the meaning of the physical meaning of $\epsilon$.} is computed as

\begin{ceqn}
\begin{equation}
 \epsilon = \frac{r_{apo} - r_{peri}}{r_{apo} + r_{peri}},
 \label{equ:eccratio}
\end{equation}
\end{ceqn}
where $r_{ apo}$ nd $r_{ peri}$ are apo- and peri-centric distances respectively. Figure \ref{fig:example} shows a visual representation of how $\epsilon$ and $\phi$ are calculated.

\begin{figure}
\centering
\includegraphics[width=0.47\textwidth]{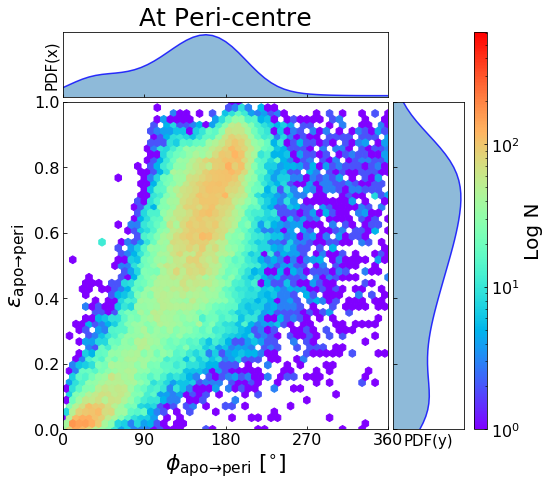}
\caption{A 2D histogram of (sub)haloes, plotting the angle subtended at peri-centric apsis as they orbit their host ($\phi$) against eccentricity orbit ($e$). The colours show log number counts and the histograms in the top and right-hand panels show the PDFs of both quantities. \label{fig:perihist}}
\end{figure}

\begin{figure}
\centering
\includegraphics[width=0.47\textwidth]{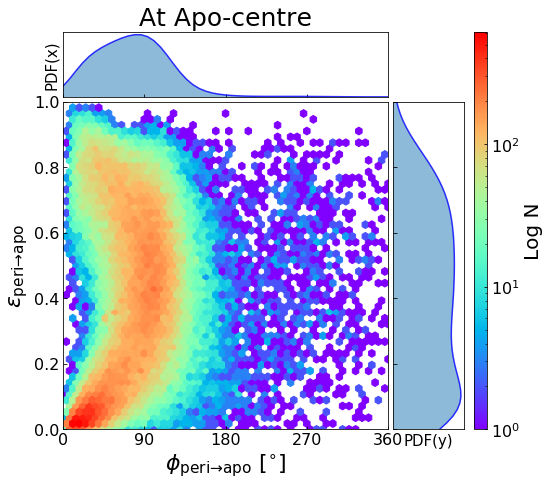}
\caption{As Figure \ref{fig:perihist}, but for apo-centric apsides.  \label{fig:apohist}}
\end{figure}

Figures \ref{fig:perihist} and ~\ref{fig:apohist} show the $\epsilon$ and $\phi$ distributions for peri-centric and apo-centric apsides respectively. These Figures show that apo-centric apsides happen at smaller $\phi$ than peri-centric apsides do, which, as we argue in \S~\ref{sec:complextoy},  occurs because most subhaloes are on in-spiralling orbits. Peri-centres, however, have a $\phi$ of close to 180$^{\circ}$ with respect to the previous apo-centre; this is because (sub)haloes experience most of their mass loss at peri-centre, with a correspondingly large change in angular momentum. This means that subhaloes transition to lower energy orbits with smaller apo-centres, which continues until the subhalo disrupts and is merged with the host.

\begin{figure}
\centering
\includegraphics[width=0.47\textwidth]{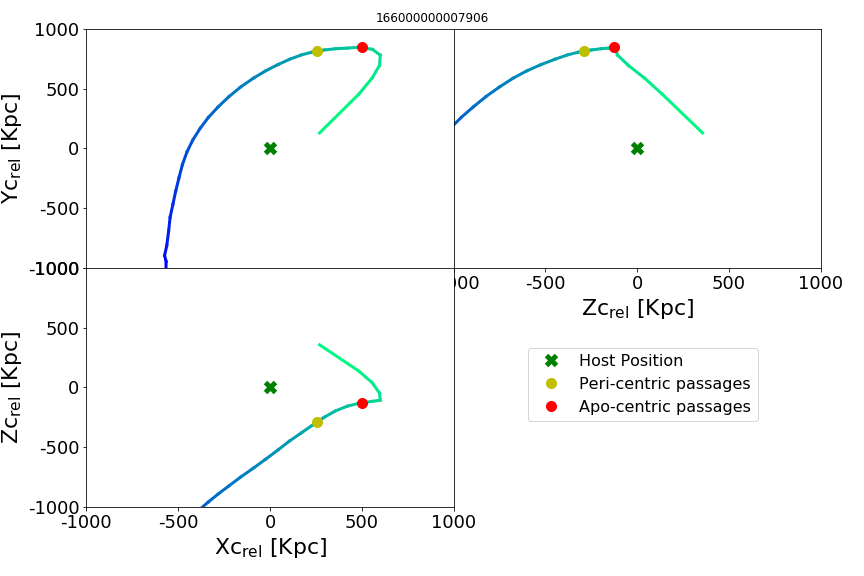}
\caption{Projections of an example orbit that undergoes a wobble as it interacts with another (sub)halo, causing it to undergo a peri-centric followed by an apo-centric apsis. \label{fig:3dorbpassage}}
\end{figure}

A striking feature of both Figure \ref{fig:perihist} and Figure \ref{fig:apohist} is that there are two populations, which can be seen clearly in the right histograms. Both Figures have a second population at low $\phi$ and $ e$. To understand this further, we plot an example orbit with $\phi$ of 17$^{\circ}$ and $e=0$; this is shown in Figure \ref{fig:3dorbpassage}, where it can be seen that these apses are due to the (sub)halo interacting with another orbiting (sub)halo of a larger mass. The interaction causes it to undergo changes in radial velocity over
a narrow radial range and to have apses close together, which results in small $\phi$ and $\epsilon$. This is the effect described previously in Section \ref{sec:cleaning} and shown in Figure \ref{fig:cleanorb}, but because it happens over a longer timescale, it is not removed by the initial clean.

From Figure \ref{fig:apohist}, it is clear that such interaction-induced low values of $\phi$ and $\epsilon$ are more likely to happen when a peri-centre is followed by an apo-centre. This is because most interactions involve a less massive subhalo orbiting its more massive host, and because the orbit of the more massive subhalo will be relatively unperturbed, it is
the less massive subhalo that will first experience an outward apsis point (V$_{ rad}$>0)
relative to the host halo, followed by an inward apsis point (V$_{ rad}$<0), thereby resulting
in repeated peri- and apo-centres. These false apsides typically happen at larger radii where
the orbiting subhalo is more likely to interact with other orbiting subhaloes.

\begin{figure}
\centering
\includegraphics[width=0.47\textwidth]{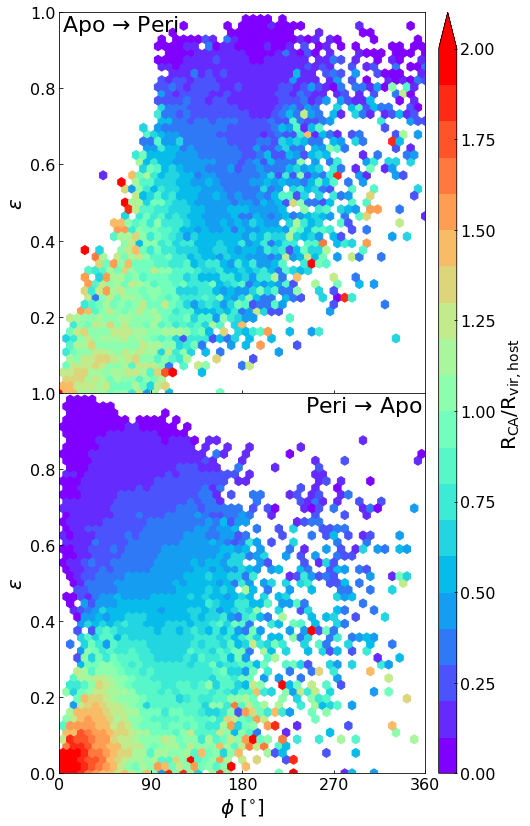}
\caption{2D histogram of distribution of $\phi$ against $e$, but colour-coded by the mean of smallest radius of closest approach, normalised to $R_{vir,host}$ in each bin, measured along the subhaloes orbit, $R_{CA}$. \label{fig:closest}}
\end{figure}

To better understand where these orbital wobbles occur, we plot a 2D histogram of $\phi$ and $\epsilon$ but colour-code by the mean smallest radius of closest approach, R$_{CA}$ (the closest the subhalo has been to its host up to this point), along each (sub)halo's orbit, normalised by R$_{vir,host}$ in each bin; this is shown in Figure \ref{fig:closest}. The top panel shows $\phi$ and $\epsilon$ for apo-centre to peri-centre; the lower panel shows peri-centre to apo-centre. In the top panel, it can be seen that orbits that tend to have a $\phi$ of 180$^{\circ}$ and high $\epsilon$ have a very small value of R$_{CA}$ at peri-centre, as expected. False apsides at low $\phi$ and $\epsilon$ tend to have a large value of R$_{CA}$ for their closest approach, typically outside R$_{vir,host}$. 

\begin{figure}
\centering
\includegraphics[width=0.47\textwidth]{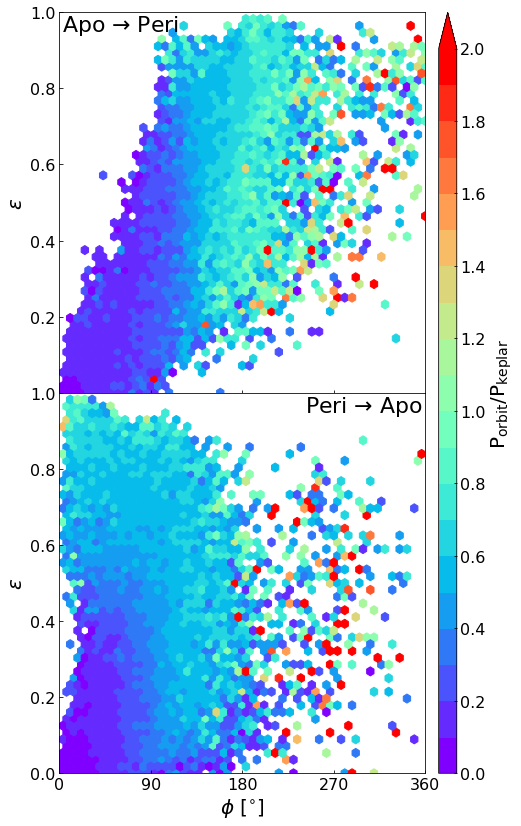}
\caption{2D histogram of distribution of $\phi$ against $e$ as in Figure \ref{fig:closest}, but 
colour-coded by the mean of the measured orbital period $P_{orbit}$ compared to the Keplerian period, calculated from Equation \ref{equ:pkepler} in each bin.   \label{fig:Pratio}}
\end{figure}

The population of orbits that {\it wobble} are clearly shown in Figure \ref{fig:Pratio}, which is the similar to Figure \ref{fig:closest} but is colour-coded by the mean ratio of measured
to Keplerian orbital periods in each bin ($P_{ orbit}$ and $P_{Kepler}$ respectively), where,
\begin{ceqn}
\begin{equation}
    P_{Kepler} = 2\pi \sqrt{\frac{a^3}{G(M_{orbit} + M_{host})}};
    \label{equ:pkepler}
\end{equation}
\end{ceqn} 
where $a$ is the semi-major axis, $M_{orbit}$ is the mass of the orbiting subhalo, and $M_{host}$ is the mass of the host halo \citep{Russell1964}. Figure \ref{fig:Pratio} clearly shows that the apsides that have low ratios of $P_{orbit}/P_{Kepler}$ are orbits with low $e$ and $\phi$. These apsides are, in fact, merely orbital wobbles, and lead to a low period because they happen over a short time, as can be seen in Figure \ref{fig:3dorbpassage}.
 
\begin{figure}
\centering
\includegraphics[width=0.47\textwidth]{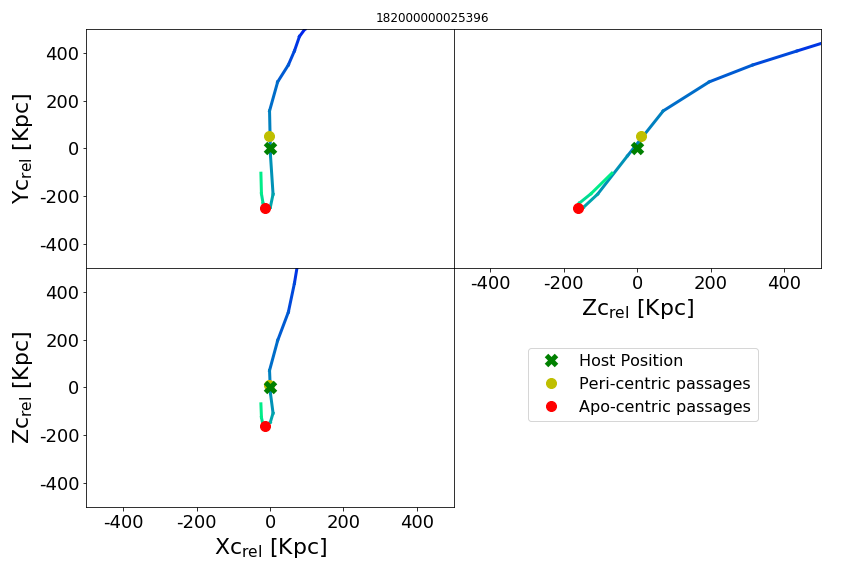}
\caption{This plot shows the projections for an example orbit which has a low $\phi$ but a high $e$, which is on a highly radial orbit. \label{fig:3dorbclosest}}
\end{figure}

From Figure \ref{fig:apohist}, there is a less apparent third population which has high $e$ but low $\phi$. This population becomes more apparent when points on the figure are colour-coded by the radius of current closest approach in each (sub)halo's orbit ($R_{CA}$) normalised to $R_{vir,host}$, as shown in Figure \ref{fig:closest}. There is a clear trend in the apo-centres, where apo-centres with low $e$ and $\phi$ tend to have large values of $R_{CA}$, while those with high $e$ and low $\phi$ tend to have smaller values of $R_{CA}$.  \\

To better understand this, we show in Figure \ref{fig:3dorbclosest} an example orbit drawn from this region. We see that the orbiting subhalo has a very radial orbit, in which it passes very close to its host and undergoes a quick change in direction of $V_{rad}$. Because it passes so close to its host, the orbiting (sub)halo loses a lot of angular momentum and is unable to maintain its orbit. This causes the (sub)halo to undergo a rapid change in velocity, and it quickly merges with its host. This apsis has a low $\phi$ because the (sub)halo could not complete a full orbit, but has a high value of $e$ since its first apsis is close to its host, but its secondary apsis happens at larger radius. 

\subsection{Toy Orbital Evolution Model} \label{sec:Toy}

To understand why there is a difference in $\phi$ at peri- and apo-centres, we model a subhalo orbiting in a NFW halo with a potential given by
\begin{ceqn}
\begin{equation}
    \rho (r)=\frac{\rho_c\delta_c}{r/R_s\left(1~+~r/R_s\right)^2},
\end{equation}
\end{ceqn}
where $r$ is the host halo-centric radius; $R_s$ = R$_{vir}$/$c$ is the scale radius, where $c$ is the concentration of the halo; $\rho_c$ is the critical density; and $\delta_c$ is the characteristic overdensity given by
\begin{ceqn}
\begin{equation}
    \delta_c = \frac{200}{3}\frac{c^3}{\ln(1~+~c)~-~c/(1~+~c)}.
\end{equation}
\end{ceqn}
We assume that the density profile is truncated just outside $R_{vir}$. The orbit is calculated in 2-dimensions over a time of 1 Gyr; a $2^{\rm nd}$ order leapfrog integrator is used with 1,000 steps equally spaced in time (i.e. 1 Myr). The initial conditions are $(X,Y)=(0,1) {\rm Mpc}$ and $(V_{X},V_{Y})=(0.5,0.1) V_{\rm vir}$, where $V_{\rm vir}$ is the virial velocity of the host halo. 

\subsubsection{A simple model} \label{sec:simpletoy}

\begin{figure}
\centering
\includegraphics[width=0.47\textwidth]{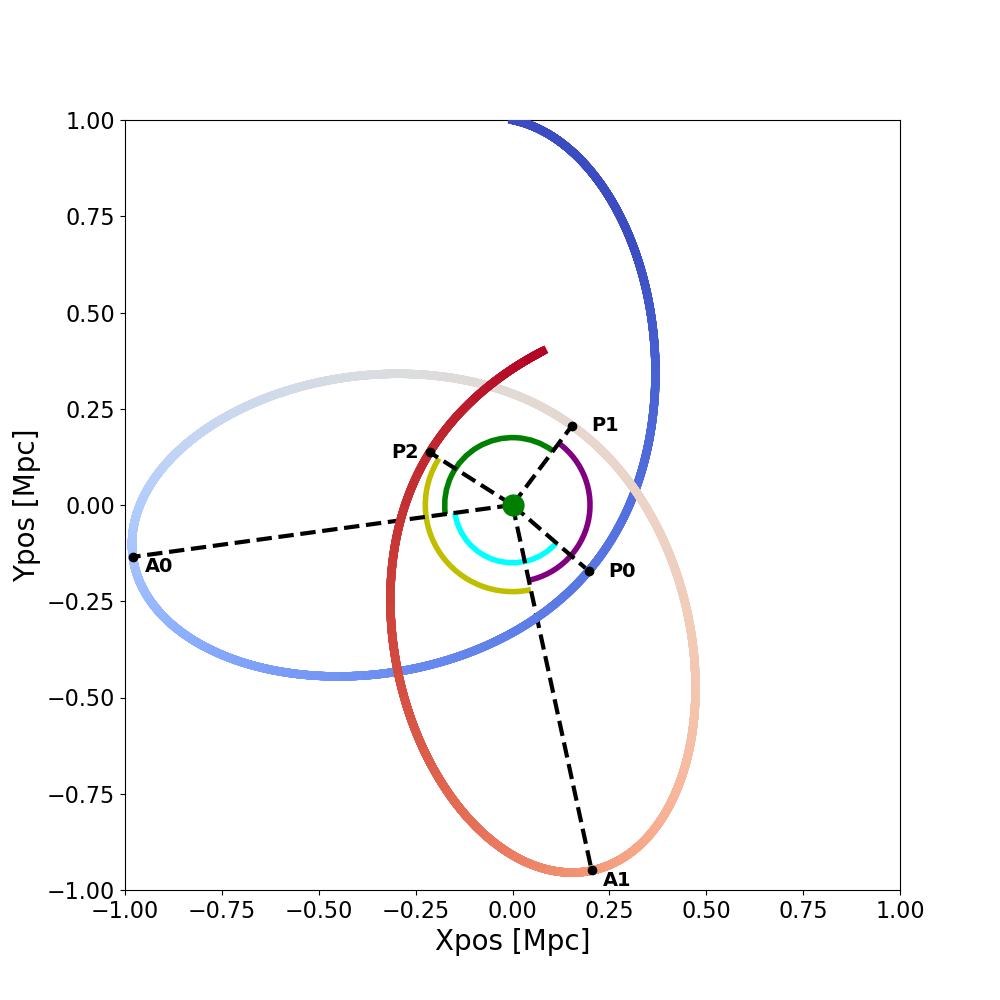}
\caption{ This figure shows the position of a point particle in a NFW potential orbiting around its host. The colour of the line represents going from low $t$ (blue) to high $t$ (red). The green point represents the host position. The letters represent the type of apsis of the orbit, where $P$ is peri-centre, and $A$ is apo-centre. The numbers represent the orbital passages, starting from `0' for the first approach. \label{fig:toy}}
\end{figure}

As our starting point, let us assume that (1) the only force acting on the subhalo is the gravity of the host halo; (2) the subhalo can be treated as a point particle with negligible relative mass; (3) the mass of the host does not change in time, and (4) its density profile is fixed; and (5) the orbital angular momentum of the subhalo is constant. The results of our model using these assumptions are shown in Figure \ref{fig:toy}. It can be seen that angles between the apsides - shown by the cyan, green, yellow and purple lines - are always below 180$^{\circ}$; in other words, an orbit in a NFW potential causes apsidal procession. In the case of the toy orbit shown in Figure \ref{fig:toy}, the angle between the apsides is a constant ~130$^{\circ}$. While this model explains why $\phi$ can be below 180$^{\circ}$, it does not explain the differences in $\phi$ found at peri-centre and apo-centre,  as seen in Figures \ref{fig:perihist} and \ref{fig:apohist}. 

\subsubsection{A more complex model} \label{sec:complextoy}
\begin{figure}
\centering
\includegraphics[width=0.47\textwidth]{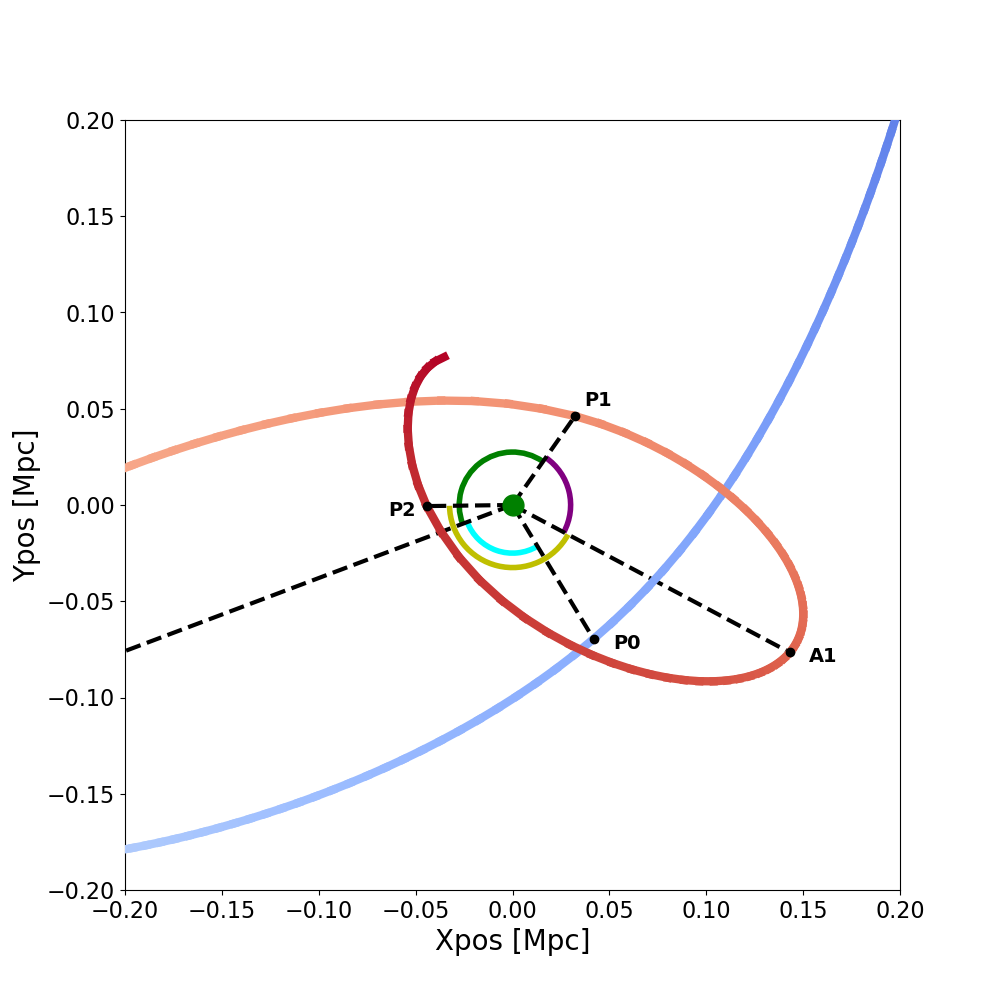}
\caption{ This figure shows the position of the point particle, but this time with the host increasing in mass and the orbiting halo losing $L_{orb}$. The labels are same as Figure \ref{fig:toy}.  \label{fig:toy2}}
\end{figure}

We now relax assumption (3), which is that the mass of the host halo is constant, and allow it to increase with time by a factor of 3 over the period of the orbit; this was found to be a typical value for mass change over 1 Gyr. We also increase the concentration of the host, $c$, by a factor of 2 over the same period. In addition, we relax assumption (5) and assume that the subhalo loses some orbital angular momentum (L$_{orb}$) at peri-centre, as would occur if the subhalo was disrupted. Results of this more complex model are shown in Figure \ref{fig:toy2}. The increase in the mass of the host and the loss of L$_{orb}$ causes the angle between peri- and apo-centre (cyan and purple angles) to be smaller than in the simple model - with fixed host mass and constant angular momentum - and close to 90$^{\circ}$. In addition, the angle between apo- and peri-center (green and yellow angles) is close to 180$^{\circ}$, which is seen in Figures \ref{fig:perihist}, \ref{fig:apohist}  \ref{fig:Pratio} and \ref{fig:closest}.  

\medskip

\noindent We now investigate which property has the most significant effect on $\phi$ -- M$_{host}$ increasing, c$_{host}$ increasing or L$_{orb}$.

\begin{figure}
\centering
\includegraphics[width=0.47\textwidth]{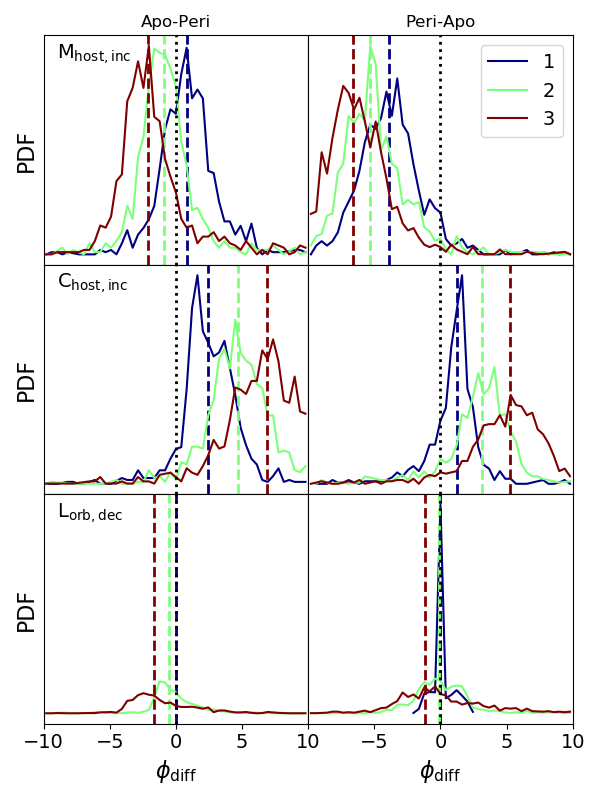}
\caption{ This plot shows the PDF of the difference in $\phi$ for the toy model with either increasing M$_{host}$ (top row), increasing C$_{host}$ (middle row) or decreasing L$_{orbt}$ (bottom row) when compared to the simple toy model in an NFW potential. The left column shows haloes going from apo-centre to peri-centre (Apo-Peri), and the right shows the opposite (Peri-Apo). The solid line shows the PDF, and the dashed line shows the median. The colour is the number of each apsis points the halo has undergone.  \label{fig:phidiff}}
\end{figure}

\subsubsection*{Increasing M$_{host}$}

Here we assess the impact of increasing of M$_{host}$ by contrasting $\phi$ in models in which M$_{host}$ is increasing with ones in which it is fixed ($\phi_{diff}$). We systematically vary the initial conditions, exploring a range of values in both models; initial positions are required to be outside a radius of 0.5 Mpc, while velocities are required to be below the escape velocity of the host at the initial position and greater than 0.3V$_{vir}$ \footnote{This ensure that the halo orbits its host and does not merge on a short timescale.}.

We plot the PDF and medians of $\phi_{diff}$ as measured when the halo has completed three peri-centric and apo-centric apsides in the upper panels of Figure \ref{fig:phidiff}; left and right panels show the cases when the subhalo passes from apo- to peri-centre (Apo-Peri) and peri- to apo-centre (Peri-Apo) respectively. Line colour indicates the number of apsides completed, where three of each apsis is shown. This shows that increasing M$_{host}$ causes $\phi$ to decrease, and for the magnitude of the decrease to increase with each apsis. This is because increasing M$_{host}$ reduces the relative orbital energy of the satellite, meaning that its orbital speed around the host is decreased and so apsis occurs more quickly. The effect is more pronounced when going from Peri-Apo, which is most likely because orbits will experience a peri-centre first, and it is post- first peri-centric passage when they lose most of their orbital energy.

\subsubsection*{Increasing c$_{host}$}

Here we assess the impact of increasing the host NFW profile concentration (c$_{host}$), following the approach used in the previous sub-Section. We keep the enclosed mass at the truncation radius fixed. The middle panels of Figure \ref{fig:phidiff} shows, as before, the PDF of the difference of this model's $\phi$ to the simple model $\phi_{diff}$. In contrast to the case of increasing M$_{host}$, we see that increasing c$_{host}$ leads to a corresponding increase in the angle when going from peri- to apo-centre ($\phi_{diff}$). This is because increasing c$_{host}$ means that M$_{encl}$ is always higher after apsis than before it. We expect that this increase will be minimal after peri-centre because this is when the orbiting halo is travelling the fastest, but it should be more pronounced after apo-centre, when $\Delta$V$_{rad}$ < 0. The subhalo's speed will increase, resulting in a shorter orbital time, and a corresponding increase in $\phi$.

\subsubsection*{Decreasing L$_{orb}$}

Finally, we assess the impact of decreasing L$_{orb}$. This decrease occurs at peri-centre, where the average L$_{orb}$ decrease is estimated from the $N$-body simulation. The PDF of $\phi_{diff}$ for this is shown in the bottom panels of Figure \ref{fig:phidiff}, where we observe that decreasing L$_{orb}$ generally leads to lower $\phi$. As with increasing $M_{host}$, the subhalo loses orbital energy, which means that it takes longer to complete an orbit, which leads to a decrease in $\phi$. However, we find that this decrease in L$_{orb}$ has a negligible effect. This may be because of the simplified treatment in our model, e.g. we neglect dynamical friction, which might be important in this regime for some subhaloes, particularly for subhaloes which are on semi-stable orbits. A more sophisticated treatment is beyond the scope of this paper.

\subsubsection*{The net effect}

Considering all the results shown in Figure \ref{fig:phidiff}, we conclude that Apo-Peri $\phi$ increases with the number of orbits because of the increase in C$_{host}$, whereas Peri-Apo $\phi$ decreases because of the increase in M$_{host}$. The overall affect is that $\phi$ increases for Apo-Peri and decrease for Peri-Apo, which is seen to happen in Figure \ref{fig:toy2}. The net effect is also what is observed to happen to the example planar orbit in Figure \ref{fig:example}, where $\phi_2$ > $\phi_1$. The model presented here offers a reasonable explanation of the effect observed.

\subsection{Orbit Cleaning}

Any automated approach to classifying orbits must account for {\it wobbles} in a subhalo's orbit. As discussed above, these may arise because of interactions with other subhaloes. After investigation, we suggest cleaning apsides that satisfy the following criteria:

\begin{ceqn}
\begin{equation}
    \epsilon \ < \ 0.4 \quad \& \quad \phi \ < \ 80.
    \label{equ:clean}
\end{equation}
\end{ceqn}

\begin{figure*}
\centering
\includegraphics[width=\textwidth]{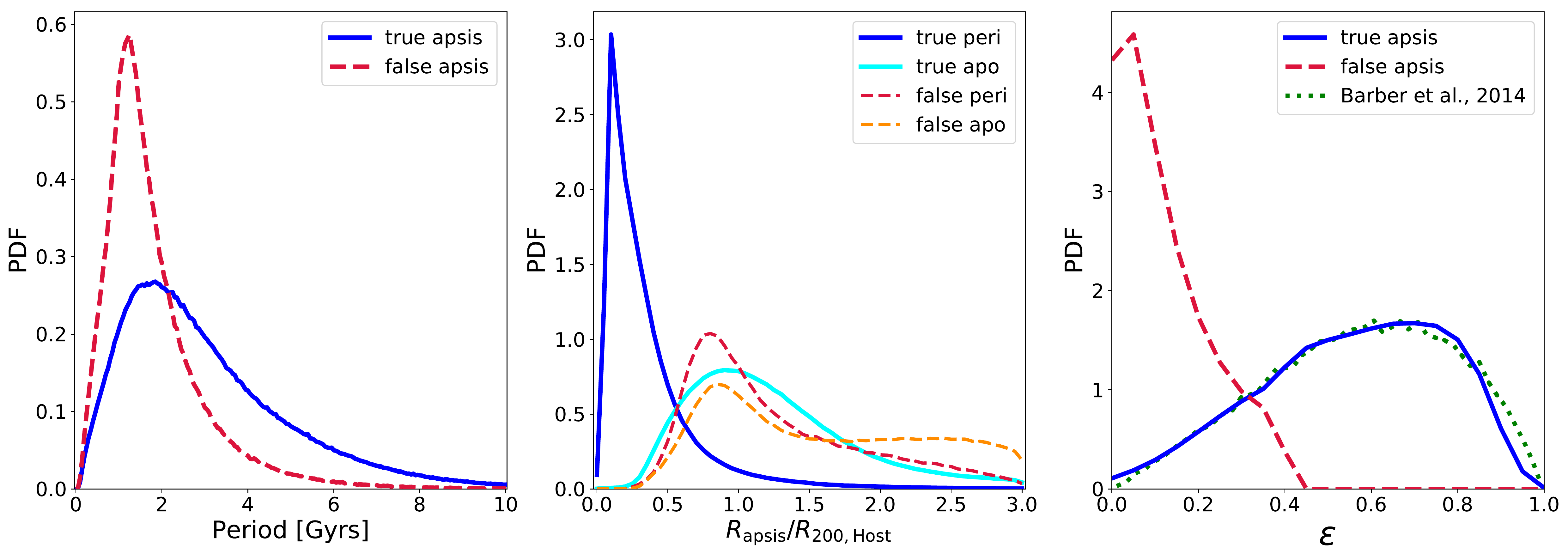}
\caption{ This figure shows the PDF of the orbital period (left panel), the radius of the apsis point relative to the R$_{vir,host}$ (middle panel), and $\epsilon$ (right panel). This is to show the difference between the clean (true) apsis points and the points which are to be cleaned (false) apsis points.  \label{fig:orbdist}}
\end{figure*}

To show the effect of removing false peri/apo-centres (i.e. those that satisfy these criteria) on the distribution of orbital periods, we plot apsis radii (R$_{apsis}$) measured relative to R$_{vir,host}$ and $\epsilon$ in Figure \ref{fig:orbdist}. We contrast the pre-cleaned distribution of peri- and apo-centres with the true, post-cleaned, and false, removed, distributions using criteria \ref{equ:clean}. This shows that false peri- and apo-centres generally have very short periods - with a mean of 1.4 Gyr - which is because apsides occur within a small time period. In contrast, the distribution of true apo- and peri-centres is much broader, with a mean of 2.8 Gyrs. The R$_{apsis}$ ratio distribution for the false peri-centres (apo-centres) peaks outside of R$_{vir,host}$ (3 R$_{vir,host}$), showing that these happen outside of the host and can be due to the subhalo orbiting another host as it comes in to merge with the host of interest. However, the defining property of false apsides is $\epsilon \sim 0$ because peri- and apo-centres occur at similar radii.

\medskip

We can see the effect of this cleaning on the orbital parameters in Figure \ref{fig:orbdist}, where we show PDFs of the orbital period; the radius of apsis (R$_{apsis}$) to R$_{vir,host}$; and $\epsilon$. We indicate pre-cleaned, true post-cleaned, and false post-cleaned PDFs by solid, dashed, and dotted curves. 

\begin{figure}
\centering
\includegraphics[width=0.47\textwidth]{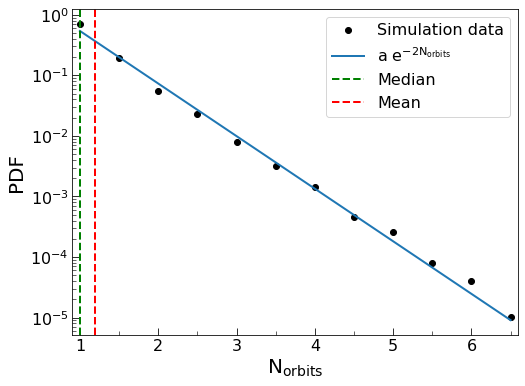}
\caption{This figure shows the projections for an example orbit which has a low $\phi$ but a high $e$, which is on a highly radial orbit. \label{fig:numorbpdf}}
\end{figure}

\begin{itemize}
\item Distribution of periods (left hand panel): we see that false apsides have a mean period of 1 Gyrs and a narrower distribution that tails off quickly at long periods, whereas true apsides have a mean period of 2 Gyrs, and a broad distribution with a long tail at long periods. 
\item Distributions of R$_{apsis}$/R$_{vir,host}$ (middle panel): we see that false peri- and apo-centres peak at radii of approximately 1 times and 2 times R$_{vir,host}$, respectively, whereas the true peri- and apo-centres peak at 0.2 and 1 R$_{vir,host}$, respectively, with 99\% of the peri-centres (apo-centres) lying within 1 (2) R$_{vir,host}$. This is in agreement with previous work \citep[e.g.][]{Khochfar2006,Ludlow2009,Barber2014} .
\item Distribution of $\epsilon$ (right hand panel): the false apsis points have a distribution that peaks sharply 
at $\epsilon$ < 0.2, as we would expect because of the small radial separation between the apsides for false orbits, whereas the distribution of true apsis points has a $\epsilon \simeq 0.82$, which is consistent with \citep{Barber2014}\footnote{This distribution is based on the orbits of surviving z=0 subhalos of Milky Way mass halos using the N-body Aquarius simulations. Differences between our results and theirs may be due to the stricter orbit selection used in \citealt{Barber2014} and the choice of MW mass hosts. However, since dark matter haloes are roughly self-similar, we expect the orbits to be roughly self-similar and difference to be small.}. 
\end{itemize}
Figure \ref{fig:numorbpdf} shows the PDF of the number of orbits that (sub)haloes experience after cleaning, where both the mean and median are shown to have a value of  approximately 1. The PDF falls off as $\sim$ e$^{\rm -2N_{orbits}}$, which suggests that most infalling (sub)haloes are on plunging radial orbits and complete only one orbit before they are disrupted. This agrees with the $\epsilon$ PDF (right-hand panel in Figure \ref{fig:orbdist}), where most orbits have a $\epsilon$ close to 1 and so are on more elliptical/ radial orbits. These type of orbits bring them close to their host, leading to a rapid mass loss. Consequently the (sub)halo is only able to survive for a single orbit.

\subsection{Effect on the Merger Timescale}

 \begin{figure*}
\centering
\includegraphics[width=0.9\textwidth]{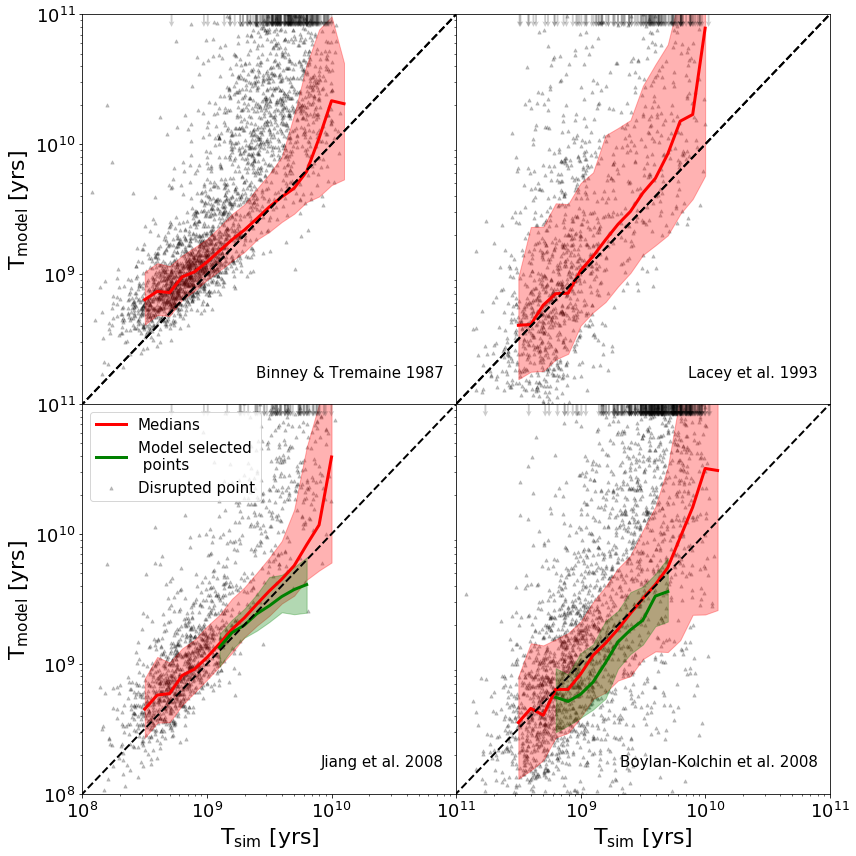}
\caption{This plot shows merger timescales measured from the simulation (T$_{sim}$) compared
to their predicted values (T$_{model}$). The model that is used is shown in the bottom right of each panel. The dashed black line shows the 1 to 1 line. The solid red line shows the medians for the whole population with the shaded region showing the standard deviation. The solid green line shows medians for the model selected point which is only for J08 (Equation \ref{equ:J08sel}) and BK08 (Equation \ref{equ:BK08sel}) and the shaded region also shows the standard deviation. The small black triangles show the points which are susceptible to artificial disruption ignored in the calculations, with the downward arrows show the points have a long T$_{model}$ than is shown on the plot. \label{fig:mtscomp}}.
\end{figure*}

The merger timescale, i.e. the time from entering R$_{vir,host}$ to the coalescence of the subhalo and the host halo, is strongly influenced by the type of orbit the subhalo is on. Many studies have looked into this effect and tried to quantify it  \citep[e.g.][]{Binney1987,lacey_merger_1993,Jiang2008,boylan-kolchin_dynamical_2008,Simha2017}. Here we assess the prescriptions presented in \citealt{Binney1987} (hereafter BT87); \citealt{lacey_merger_1993} (hereafter L93); \citealt{Jiang2008} (hereafter J08); and \citealt{boylan-kolchin_dynamical_2008} (hereafter BK08), and compare these prescriptions' predictions with measurements from $N$-body simulations in Figure \ref{fig:mtscomp}. The panels show the measured merger timescales plotted against the merger timescales predicted by BT87 and L93 (upper left and right) and by J08 and BK08 (lower left and right). The red curves and shaded regions indicate medians and standard deviations for all of the haloes in our simulations.

Note that because the J08 and BK08 prescriptions are derived from $N$-body simulations, we take care to apply the
same selection criteria to our sample of haloes; the J08 criteria are:

\begin{ceqn}
\begin{equation}
\frac{M_{sat}}{M_{host}} > 0.065 \quad{} \& \quad{} \frac{R_c}{R_{vir,host}} < 1.5 \quad{} \& \quad{} Z > 2.
\label{equ:J08sel}
\end{equation}
\end{ceqn}

While the BK08 criteria are:

\begin{ceqn}
\begin{equation}
\frac{M_{sat}}{M_{host}} > 0.025 \quad{} \& \quad{} 0.65<\frac{R_c}{R_{vir,host}} < 1.0 \quad{} \& \quad{} 0.3 < \eta < 1.0.
\label{equ:BK08sel}
\end{equation}
\end{ceqn}

These criteria define the subset of haloes for which medians (green curves) and standard deviations (shaded regions) are estimated in the lower panels of Figure \ref{fig:mtscomp}.

We highlight the subset of (sub)haloes that are likely to have undergone artificial disruption (black triangles), which we exclude when calculating medians and standard deviations. The primary cause of artificial disruption is a combination of insufficient mass resolution and force softening, and this causes overmerging - the dissolution of subhaloes on a more rapid timescale than would be expected, given the physical conditions \citep[e.g. the gravitational tidal field of the host][]{vandenBosch2018,vandenBosch2018a}. This means that a subset of the subhaloes that have merged in the simulation may not merge if simulated at a higher resolution. To remove haloes that we believe have undergone artificial disruption, we impose two requirements:

\begin{itemize}
    \item a particle number cut such that all subhaloes must have in excess of 1,000 particles at infall; and 
    \item the ratio of the minimum scale radius ($R_{s,min}$) along the subhaloes orbit is always above 2 times the force softening length (see appendix \ref{app:disrupt}). 
\end{itemize}   

\medskip

These results indicate that BT87 is a reasonable approximation for the merger time; for shorter merger times ($\sim 10^9$ yr), the median tends to be over-predicted, which is to be expected because the prescription does not account for changes in the orbit that arise over the merger timescale (e.g. \citealt{lacey_merger_1993,Jiang2008,boylan-kolchin_dynamical_2008}). If we consider only the sample of subhaloes that satisfy the J08 and BK08 criteria, then the median behaviour is such that merger times are under-predicted slightly.  

All of the prescriptions over-predict the merger times of those subhaloes with the largest measured values (corresponding to the largest host-to-subhalo mass ratios), suggesting that they will not merge over the lifetime of the simulation. Most of these objects have undergone a violent event in their past and have suffered significant mass loss, much higher than predicted if it is driven by gradual stripping by the smooth background; this causes them to merge much earlier than predicted in practice. The subset of subhaloes that are predicted to undergo artificial disruption follow the same trend as the primary sample, but show substantial scatter. This is because these systems undergo early disruption, which depends on the type of orbit they are on and influences where they end up on the plot.

 \begin{figure}
\centering
\includegraphics[width=0.45\textwidth]{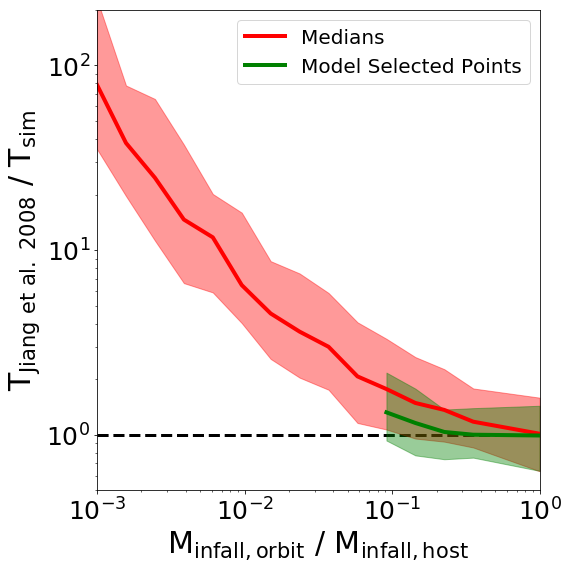}
\caption{This figure shows the ratio of the satellite's mass to the host mass at infall against the ratio of the predicted merger timescale to the simulated merger timescales. The red line shows the medians of all the satellites in the simulation and the red shaded region shows the standard deviation around the median. The green line shows the selected haloes from Equation \ref{equ:J08sel}, and the green shaded region shows the standard deviation of the sample. The black dashed line shows when the predicted and simulated merger timescales agree.}
\label{fig:MassRatioVsMTS}
\end{figure}

To understand this further, we have taken the best fitting model J08 and plotted the ratio of the satellites to host mass at infall against the ratio of the predicted merger timescale to the simulated merger timescale. This is shown in Figure \ref{fig:MassRatioVsMTS}, where the red line shows the median in each mass bin and the shaded region shows the standard deviation. The plot shows that J08 can correctly predict the life of the for the model selected points, with the medians shown in the green and the shaded region showing the standard deviation about the median. However, it over-predicts the lifetime for all smaller mass satellites relative to their host by up to 100 times.  This is most likely because the prescription does not include the effects of the tidal field. Smaller satellites are highly susceptible to this, meaning they can be catastrophically tidally disrupted.

\section{Discussion \& Conclusions}

We have introduced two new tools to study the orbital histories of subhaloes - \WW\ and \ORB\. \WW\ uses halo group catalogues and merger trees to ensure that haloes and subhaloes are tracked accurately in dense environments, while \ORB\ quantifies each (sub)halo's orbital parameters. \WW\ enhances the information that can be extracted from an $N$-body simulation, and in the process, it can change a (sub)halo's evolutionary history; for example, subhaloes that appear to have merged may have survived and exited the host halo. The effect of adding \WW\ (sub)haloes has enabled a more detailed study of (sub)halo orbits by \ORB\ because they can be tracked for longer and into higher density regions before they physically merge with their host.

\ORB\ provides a fast and efficient way to extract orbits from a large $N$-body simulations. This enables sampling of a broad range of orbits probing very different environments, which facilitates studies into how the environment can affect the type of orbits that are present and to characterise the types of interactions present (e.g.\ see Bakels et al., in prep). In addition, it is also possible to use orbital information to find where stripped material is deposited in its host \citep{Canas2019}. This will be the focus of future works using the outputs of \ORB\ run on the SURFS and Genesis suite of simulations.

In this work we have shown that results from \ORB\ are in good agreement with the literature. We also presented orbits on the $\epsilon$ (eccentricity) - $\phi$ (angle subtended since last apsis) plane. These plots show that typical apo-centric apsides tend to have lower $\phi$ and $\epsilon$ than peri-centric apsides. We demonstrated, using a toy model that simulates satellite orbits in an evolving NFW potential, that the major cause of orbital evolution is the mass growth of the parent halo. 

We also found that there is a dominant secondary population present at low $\epsilon$ and $\phi$ that arise from orbital {\it wobbles} - these are found at large halo-centric radii and have short periods, which means that they can be cleaned efficiently from an orbital catalogue by applying cuts in $\epsilon$ and $\phi$. These cuts in $\epsilon$ and $\phi$ are shown to remove most of the wobbles and can reproduce orbital distributions recovered in previous works. This improves the accuracy of the orbital data, giving the correct orbital histories of the haloes around their hosts.

Finally, we assessed how well-established prescriptions for the merger timescale compare to those measured in the simulation. We found that the prescriptions are reasonable approximations for systems with short to intermediate merging timescales, but they all over-predict the merging timescales of systems with the longest measured times. This can have profound effects on the satellite populations, since many of the satellites which should have merged in the simulation `survive' because of the incorrect lifetime given by merger timescale formulae. This is important in SAMs, since in coarser resolution simulations merger timescale prescriptions are the main mechanism used to merge subhaloes into their parent haloes \citep{robotham_galaxy_2011}.

The literature merger timescale prescriptions that were empirically recovered (J08 and BK08) only correctly predict the lifetimes within their orbital selection regimes, but over-predict the lifetime for some satellites by over 100 times if they exist outside of the selected regime. We will follow this up in future work, with the aim to better understand what is causing satellites with longer merger timescales to merge earlier than predicted by previous analytic and numeric prescriptions.

\section*{Acknowledgements}

We would like to thank Ainulnabilah B. Nasirudin and the anonymous referee for their clear and constructive comments. We would also like to thank Lucie Bakels for the helpful comments and discussions. RP is supported by a University of Western Australia Scholarship. PJE is supported by the ARC Centre of Excellence ASTRO 3D through project number CE170100013 Part of this research was undertaken on Raijin, the NCI National Facility in Canberra, Australia, which is supported by the Australian commonwealth Government. Parts of this research were conducted by the Australian Research Council Centre of Excellence for All Sky Astrophysics in 3 Dimensions (ASTRO 3D), through project number CE170100013. This research was undertaken with the assistance of resources from the National Computational Infrastructure (NCI Australia), an NCRIS enabled capability supported by the Australian Government. \\
\vspace{0.05cm}

\noindent
\textit{Software}
\begin{itemize}
\item \textsc{VELOCIraptor}: https://github.com/pelahi/VELOCIraptor-STF
\item \textsc{TreeFrog}: https://github.com/pelahi/TreeFrog
\item \textsc{VELOCIraptor_Python_Tools}: https://github.com/pelahi/VELOCIraptor\_Python\_Tools 
\item \textsc{WhereWolf}: https://github.com/rhyspoulton/WhereWolf
\item \textsc{OrbWeaver}: https://github.com/rhyspoulton/OrbWeaver
\end{itemize}

\noindent
\textit{Additional software}: Python, Matplotlib \citep{Hunter2007}, Numpy \citep{vanderWalt2011}, scipy \citep{Jones2001SciPy.org} and \textsc{Gadget} \citep{springel_simulations_2005}.




\bibliographystyle{mnras}
\interlinepenalty=10000 
\bibliography{references}  



\appendix
\section{How OrbWeaver works} \label{app:orb}

\ORB\ utilises a python script which uses the halo catalogues and a merger tree as input to generate a pre-processed file containing all the orbiting objects for a host. The user can supply selection criteria to extract the type of host desired. For this paper, the haloes of interest are set to be host haloes that had over 50,000 particles at some point in their history. 

Once a halo of interest is found, then all haloes that come within N*R$_{ vir}$ (N is user-definable) of this halo across its full existence (the branch of interest) are extracted. To be included in the "orbital forest" for the host, it must be smaller than the host and be a halo (not a subhalo) when it is first found. The orbital forests are output to a pre-processed file containing N$_{forest}$ per file (user-definable).

The \ORB\ {\sc C}-code can then be run on those pre-processed files, where each orbiting halo is given a unique orbitID to identify its orbit in the simulation.  \ORB\ can be run concurrently on each of the files. It will follow the full history for each halo, first interpolating any points where the halo is missing. Cubic interpolation is computed for position and velocity values, and logarithmic interpolation for mass, radius and V$_{max}$. 

The next step is to calculate the orbital properties for each orbiting halo, where interpolation is done at the apsis points and crossing points relative to its host. Once all the points are found, then orbit cleaning is performed (if desired), to remove any false apsis points in the haloes orbit. 

The orbit properties are output to a file containing datasets of the orbit properties each with length of the total number of apsis + crossing points for all haloes processed by \ORB\. One of these datasets is called entrytype, which can be used to identify the type of entry at this index in the datasets. The values present in this dataset are:

\begin{itemize}
\item[] -99 = Apo-center, 
\item[] 99 = pericenter, 
\item[] 0 = endpoint of an orbit (the halo has either; terminated or merged(see  the MergedFlag dataset) with another halo, merged with its host  or host has terminated or merged (see the hostMergedFlag dataset) ) 
\item[] All other values = the fraction of host crossing, i.e. entrytype * R$_{host}$ crossing (positive if infalling and negative if outfalling). To get the correct number of R$_{host}$, this dataset will need to be rounded to the desired number of decimals.
\end{itemize}

The num\_entrytype dataset can be used to identify the number which this entry is (i.e. first pericentre). To find the apsis and crossing points belonging to the same orbit, the OrbitID dataset can be used to find the entries with the same OrbitID.

\section{Removal of haloes undergoing artificial disruption} \label{app:disrupt}

\begin{center}
\includegraphics[width=0.45\textwidth]{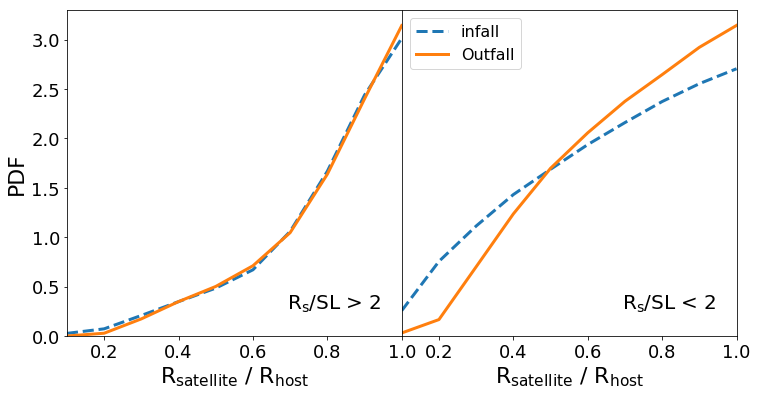}
\captionof{figure}{This figure shows the radial profile of satellites relative to their host}\label{fig:artdisrupt}
\end{center}

To show that we are excluding haloes that have possibly undergone artifical disruption, we plot the radial distribution of satellites within  R$_{vir,host}$. We split the satellite population into those that are infalling (blue dashed line) and those which are outgoing (solid). These are shown for different values of the subhaloes scale radius (R$_{s}$) to the simulation softening length (SL). The left panel shows this for satellites with R$_{s}$/SL > 2 and the right panel shows for R$_{s}$/SL < 2. The left panel shows that the population of infalling and outgoing are the same but the panel in right shows a suppression for the outgoing in the inner regions. This is suggesting that the objects with R$_{s}$/SL < 2 are being disrupted in the inner regions, so we correct for this by ensuring that the subhaloes we use always have R$_{s}$/SL > 2.

\section{Different definitions of eccentricity} \label{app:compEccDef}

\begin{center}
\includegraphics[width=0.45\textwidth]{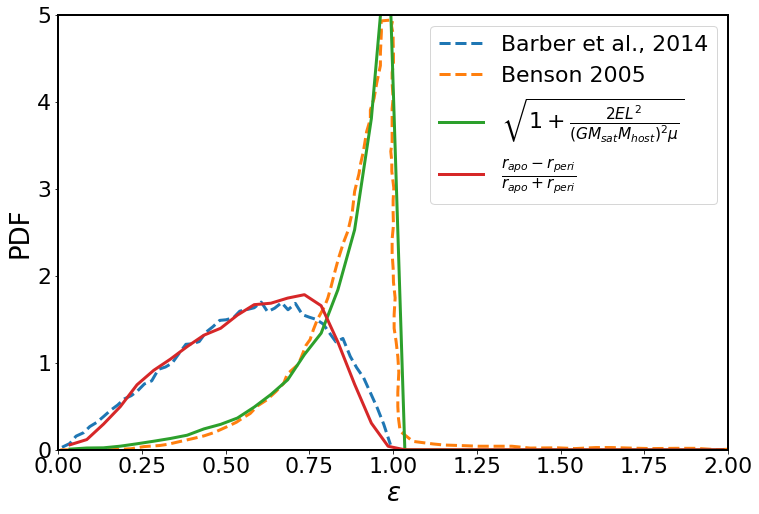}
\captionof{figure}{The distributions of the different definitions of $\epsilon$}\label{fig:compEccDef}
\end{center}

We note that in the literature there are different definitions of $\epsilon$ with the first one being in equation \ref{equ:eccratio}  and another one being (\S 2 of \citealt{Binney1987}):
\begin{ceqn}
\begin{equation}
    \epsilon \ = \ \sqrt{1 + \frac{2EL^2}{(GM_{orbit}M_{host})^2\mu}}
\end{equation}
\end{ceqn}
where, $E$ and $L$ are the orbital energy and angular momentum, with reduced mass $\mu = M_{sat}M_{host}/(M_{sat} + M_{host})$. In Figure \ref{fig:compEccDef} we compare these two different definitions and show how our simulation matches up with published results for each of these definitions, even though these are described as the same quantity.

\clearpage


\section*{Supplementary material}

## APPENDIX D: ORBWEAVER OUTPUT FIELDS

| Field | Units | Output point (Apsis or Crossing point or Both) | Description |
|---|---|---|---|
| OrbitID | N/A | Both | Unique ID to identify this orbit in each file |
| HaloID_orbweaver | N/A | Both | Unique ID to identify this halo in the pre-processed catalog in each file. The 0 values represent interpolated halos that do not exist in the pre-processed catalog |
| HaloID_orig | N/A | Both | Unique ID to identify this halo in the orignal catlogue. 0 values means it is a interpolated halo that does not exist in the original catalog |
| HaloRootProgen_orig | N/A | Both | Unique ID to indentify the root progenitor for the halo from the original halo catlog |
| HaloRootDescen_orig | N/A | Both | Unique ID to indentify the root descendant for the halo from the original halo catlog |
| OrbitedHaloID_orig | N/A | Both | Unique ID to identify the orbit host halo in the orginal catalogue. 0 values means it is a interpolated halo that does not exist in the original catalog |
| OrbitedHaloRootProgen_orig | N/A | Both | Unique ID to identify the original root progenitor of the orbited halo in the halo catalog. This can also be used to find any halos that have orbited this object, by finding the halos that have the same OrbitedHaloRootProgen_orig ID. |
| OrbitedHaloRootDescen_orig | N/A | Both | Unique ID to identify the original root descendant of the orbited halo in the halo catalog. This can be used to extract the super set of halo that orbit this host and also the halos that orbit any object that merges with this host, by finding the halos that have the same OrbitedHaloRootDescen_orig ID. |
| entrytype | N/A | Both | This value states if this entry is either: **-99** = Apo-center, **99** = pericenter, **0** = endpoint of a orbit (the halo has either; terminated/merged with another halo, merged with its host [see MergedFlag] or host has terminated/ merged [see hostMergedFlag] ) **All other values** = the fraction of host crossing i.e. entrytype * R_200crit_host crossing (positive if infalling and negative if outfalling). To get the correct number of R_200crit_host, this dataset will need to be rounded to the desired number of decimals. |
| num_entrytype | N/A | Both | This values tells you the number of each entry type so far in the orbit, such that if you want to extract the first crossing of rvir you can query the whole dataset using: entrytype==1.0 & num_entrytype==1 |
| numorbits | N/A | Both | Number of orbits the halo has completed since its first pericentric passage |
| totnumorbits | N/A | Both | The total number of orbits that the halo has completed in the simulation |
| orbitalperiod | Gyr | Apsis | Current period of its orbit |
| orbitecc_ratio | N/A | Apsis | The orbital eccetricity found from the peri/apo-centric distances in the |

| | | | simulation |
|---|---|---|---|
| orbitalenergy_inst | solarmasses km^2 / s^2 | Both | The instantaneous energy of the orbit |
| orbitalenergy_ave | solarmasses km^2 / s^2 | Apsis | The average energy of the orbit since infall or last passage, only outputted at apsis points |
| R_circ | phys Mpc | Both | The radius of a circular orbit with the same orbital energy, calculated from Khochfar and Burkert 2006 |
| V_circ | km/s | Both | The velocity of a circular orbit with the same orbital energy, calculated from Khochfar and Burkert 2006 |
| L_circ | solar masses phys Mpc km/s | Both | The orbital angular momentum of a circular orbit with the same orbital energy, calculated from Khochfar and Burkert 2006 |
| Eta | N/A | Both | The ratio of the (instantaneous) orbital angular momentum to the orbital angularmomentum of a circular orbit with the same orbital energy (L_circ). This is useful to identify the type of orbit the halo is on, where 0 is a highly radial orbit and 1 is a circular orbit. L_circ is calculated from Khochfar and Burkert 2006 |
| Rperi_calc | phys Mpc | Both | The calculated peri-centric distance from Wetzel, 2011. |
| Rapo_calc | phys Mpc | Both | The calculated apo-centric distance from Wetzel, 2011. |
| orbitecc_calc | N/A | Both | The calculated eccentricity of its orbit from Khochfar and Burkert 2006. |
| closestapproach | phys Mpc | Both | Closest approach the halo has had to is host up to this point in its orbit |
| closestapproachscalefactor | N/A | Both | The scalefactor which the closest approach occured at |
| masslossrate_inst | solar masses/Gyr | Both | The instantaneous rate at which the halo is losing mass (negative values means mass has been accreted) |
| masslossrate_ave | solar masses/Gyr | Apsis | The average rate at which the halo is losing mass, this is only calculated at apsis points so will be average since its last passage (negative values means mass has been accreted) |
| LongAscNode | Radian | Apsis | The angle of longitude of the ascending node with respect to the intial orbital plane defined here. |
| Inclination | Radian | Apsis | The inclination of the halos orbit with respect to the intial orbital plane |
| ArgPeriap | Radian | Apsis | The argument of periapsis with respect to the intial orbital plane |
| HostAlignment | Radian | Apsis | The alignment of the orbital angular momentum vector with the host halo's angular momentum vector |
| Phi | Radian | Apsis | The angle moved through since last passage |
| scalefactor | N/A | Both | Scalefactor of this entry |
| uniage | Gyr | Both | Age of the universe of this entry |
| X | phys Mpc | Both | X position of the halo in the simulation |
| Y | phys Mpc | Both | Y position of the halo in the simulation |
| Z | phys Mpc | Both | Z position of the halo in the simulation |

| VX | km/s | Both | X component of the halos velocity in the simulation |
|---|---|---|---|
| VY | km/s | Both | Y component of the halos velocity in the simulation |
| VZ | km/s | Both | Z component of the halos velocity in the simulation |
| npart | N/A | Both | Number of particles in the orbiting halo |
| Mass | solar masses | Both | Mass of the halo (depends on mass definition given) |
| Radius | phys Mpc | Both | Radius of the halo (depends on mass definition given) |
| min_Rscale | phys Mpc | Both | Minimum of the scale radius in this orbit history, used to see if the halo can be disrupted artificially due to inadequate force softening |
| min_Rmax | phys Mpc | Both | Minimum of the radius that the maximu circular velocity is at in this orbit history, used to see if the halo can be disrupted artificially due to inadequate force softening. |
| Rmax | phys Mpc | Both | Radial distance of Vmax |
| Vmax | km/s | Both | Maximum circular velocity of the halo |
| Vmaxpeak | km/s | Both | The peak Vmax has had in its existence up to the current entry time |
| cNFW | N/A | Both | Concentration of the halo |
| fieldHalo | N/A | Both | Flag if this orbiting halo is top of its spatial hierachy (not a subhalo) ,where: 0 = No, 1 = Yes |
| numSubStruct | N/A | Both | The number of substructure that this halo contains |
| RatioOfMassinSubsStruct | N/A | Both | The ratio of how much this halo's mass is in substructure |
| MergerTimeScale | Gyr | Crossing Point (see description) | How long the halo takes to merge once crossing 1.0 Rvir of its host halo, this is set the first time the orbiting halo crosses 1.0 Rvir (entrytype==1.0 & num_entrytype==1) |
| MergedFlag | N/A | Both | A flag identifying if this halo merges with its orbit host |
| Vrad | km/s | Both | The radial velocity of the orbiting halo with respect to its host |
| Vtan | km/s | Both | The tangential velocity of the orbiting halo with respect to its host |
| R | phys Mpc | Both | The distance of the halo relative to its host |
| Xrel | phys Mpc | Both | X position of the halo relative to its host |
| Yrel | phys Mpc | Both | Y position of the halo relative to its host |
| Zrel | phys Mpc | Both | Z position of the halo relative to its host |
| Vrel | km/s | Both | The halos velocity relative to its host |
| VXrel | km/s | Both | X component of the halos velocity relative to its host |
| VYrel | km/s | Both | Y component of the halos velocity relative to its host |
| Vzrel | km/s | Both | Z component of the halos velocity relative to its host |
| Lrel_inst | solar masses phys Mpc km/s | Both | The instantaneous orbital angular momentum vector |

| LXrel_inst | solar masses phys Mpc km/s | Both | The instantaneous x component of the orbital angular momentum vector |
|---|---|---|---|
| LYrel_inst | solar masses phys Mpc km/s | Both | The instantaneous y component of the orbital angular momentum vector |
| LZrel_inst | solar masses phys Mpc km/s | Both | The instantaneous z component of the orbital angular momentum vector |
| Lrel_ave | solar masses phys Mpc km/s | Apsis | The average orbital angular momentum vector since last apsis point |
| LXrel_ave | solar masses phys Mpc km/s | Apsis | The average x component of the orbital angular momentum vector since last apsis point |
| LYrel_ave | solar masses phys pc km/s | Apsis | The average y component of the orbital angular momentum vector since last apsis point |
| LZrel_ave | solar masses phys Mpc km/s | Apsis | The average z component of the orbital angular momentum vector since last apsis point |
| Npart_host | N/A | Both | Number of particle in its host halo |
| Mass_host | solar masses | Both | Mass of its host (depends on mass definition given) |
| Radius_host | phys Mpc | Both | Radius of its host (depends on mass definition given) |
| Rmax_host | phys Mpc | Both | Radial distance of Vmax for its host |
| Vmax_host | km/s | Both | Maximum circular velocity of its host |
| cNFW_host | N/A | Both | Concentration of its host |
| fieldhalo_host | N/A | Both | If this halo host is the top of its spatial hierachy (not a subhalo), where: 0 = No, 1 = Yes |
| hostMerges | N/A | Both | A flag to indicate if the orbit host merges with another halo or not where 1 = merges, 0 = does not merge |
| numSubStruct_host | N/A | Both | The number of substructure that its host contains |
| RatioOfMassinSubsStruct_host | N/A | Both | The ratio of how much of its host's mass is in substructure |




\bsp	
\label{lastpage}
\end{document}